\documentclass[11pt]{article}

\usepackage{epsfig}
\usepackage{latexsym}
\usepackage{url}
\usepackage[hypertexnames=false,hyperfootnotes=false]{hyperref}
\usepackage{texnansi}
\usepackage{color}
\usepackage{afterpage}
\usepackage{enumerate}
\usepackage{algorithm}
\usepackage{algorithmic}
\usepackage{float}
\usepackage[normalem]{ulem}
\usepackage{lmodern}
\usepackage{amsmath}
\usepackage{amsthm}
\usepackage{amssymb}
\usepackage{amsfonts}
\usepackage{comment}
\usepackage{epstopdf}
\usepackage[margin=10pt,font=small,labelfont=bf]{caption}
\usepackage{natbib}
\usepackage[margin=1in]{geometry}
\usepackage{setspace}
\newcommand{\field}[1]{\ensuremath{\mathbb{#1}}}	\newcommand{\N}{\ensuremath{\field{N}}} 
\newcommand{\R}{\ensuremath{\field{R}}} 
\newcommand{\1}{\ensuremath{\mathbf{1}}} 
\newcommand{\E}{\ensuremath{\mathsf{E}}} 
\newcommand{\defeq}{\ensuremath{\triangleq}}

\DeclareMathOperator*{\argmin}{argmin}

\newtheoremstyle{thm-sf}{}{}{\itshape}{}{\sffamily\bfseries}{.}{ }{}
\theoremstyle{thm-sf}

\newtheorem{assumption}{Assumption}

\newtheorem{theorem}{Theorem}

\newtheorem{corollary}{Corollary}
\newtheorem{lemma}{Lemma}

\newtheorem{proposition}{Proposition}

\floatstyle{ruled}
\newcommand{\emailhref}[1]{\href{mailto:#1}{\tt #1}} 

\allowdisplaybreaks[4]

\title{Adaptive Execution: Exploration and Learning of Price Impact}

\author{
Beomsoo Park \\
Electrical Engineering \\
Stanford University \\
\emailhref{beomsoo@stanford.edu} \\
\and
Benjamin Van Roy \\
Management Science \& Engineering \\
Electrical Engineering \\
Stanford University \\ 
\emailhref{bvr@stanford.edu}\\
}

\begin{document}

\maketitle

\doublespacing

\begin{abstract}
We consider a model in which a trader aims to maximize expected risk-adjusted profit while trading a single security.  In our model, each price change is a linear combination of observed factors, impact resulting from the trader's current and prior activity, and unpredictable random effects.  The trader must learn coefficients of a price impact model while trading.  We propose a new method for simultaneous execution and learning -- the confidence-triggered regularized adaptive certainty equivalent (CTRACE) policy -- and establish a poly-logarithmic finite-time expected regret bound.  This bound implies that CTRACE is {\it efficient} in the sense that the $(\epsilon,\delta)$-convergence time is bounded by a polynomial function of $1/\epsilon$ and $\log(1/\delta)$ with high probability.  In addition, we demonstrate via Monte Carlo simulation that CTRACE outperforms the certainty equivalent policy and a recently proposed reinforcement learning algorithm that is designed to explore efficiently in linear-quadratic control problems.
\end{abstract}

{\it Key words}: adaptive execution, price impact, reinforcement learning, regret bound

\section{Introduction}

A large block trade tends to ``move the market'' considerably during its execution by either disturbing the balance between supply and demand or adjusting other market participants' valuations.  Such a trade is typically executed through a sequence of orders, each of which pushes price in an adverse direction.  This effect is called {\it price impact}.  Because it is responsible for a large fraction of transaction costs, it is important to design execution strategies that effectively manage price impact.  In light of this, academics and practitioners have devoted significant attention to the topic [\cite{BertsimasLo98,AlmgrenChriss00,KissellGlantz03,ObizhaevaWang05,MoallemiParkVanRoy08,AlfonsiSchiedSchulz07b}]. 

The learning of a price impact model poses a challenging problem.  Price impact represents an aggregation of numerous market participants' interpretations of and reactions to executed trades.  As such, learning requires ``excitation'' of the market, which can be induced by regular trading activity or trades deliberately designed to facilitate learning.  The trader must balance the short term costs of accelerated learning against the long term benefits of an accurate model.  Further, given the continual evolution of trading venues and population of market participants, price impact models require retuning over time. In this paper, we develop an algorithm that learns a price impact model while guiding trading decisions using the model being learned.  

Our problem can be viewed as a special case of reinforcement learning.  This topic more broadly addresses sequential decision problems in which unknown properties of an environment must be learned in the course of operation (see, e.g., \cite{SuttonBarto98}).  Research in this area has established how judicious investments in decisions that explore the environment at the expense of suboptimal short-term behavior can greatly improve longer-term performance.  What we develop in this paper can be viewed as a reinforcement learning algorithm; the workings of price impact are unknown, and exploration facilitates learning. 

In reinforcement learning, one seeks to optimize the balance between exploration and exploitation -- the use of what has already been learned to maximize rewards without regard to further learning.  Certainty equivalent control (CE) represents one extreme where at any time, current point estimates are assumed to be correct and actions are made accordingly.  This is an instance of pure exploitation; though learning does progress with observations made as the system evolves, decisions are not deliberately oriented to enhance learning. 

An important question is how aggressively a trader should explore to learn a price impact model.   Unlike many other reinforcement learning problems, in ours a considerable degree of exploration is naturally induced by exploitative decisions.  This is because a trader excites the market through regular trading activity regardless of whether or not she aims to learn a price impact model.  This activity could, for example, be triggered by return-predictive factors, and given sufficiently large factor variability, the induced exploration might adequately resolve uncertainties about price impact.  Results of this paper demonstrate that executing trades to explore beyond what would naturally occur through exploitation can yield significant benefit.  

Our work is constructive: we propose the {\it confidence-triggered regularized adaptive certainty equivant} policy (CTRACE), pronounced ``see-trace,'' a new method that explores and learns a price impact model alongside trading.  CTRACE can be viewed as a generalization of CE, which at each point in time estimates coefficients of a price impact model via least-squares regression using available data and makes decisions that optimize trading under an assumption that the estimated model is correct and will be used to guide all future decisions.  CTRACE deviates in two ways: (1) $\ell_2$ regularization is applied in least-squares regression and (2) coefficients are only updated when a certain measure of confidence exceeds a pre-specified threshold and a minimum inter-update time has elapsed. Note that CTRACE reduces to CE as the regularization penalty, the threshold, and the minimum inter-update time vanish. 

We demonstrate through Monte Carlo simulation that CTRACE outperforms CE.  Further, we establish a finite-time regret bound for CTRACE; no such bound is available for CE.  {\it Regret} is defined here to be the difference between realized risk-adjusted profit of a policy in question and one that is optimal with respect to the true price impact model.  Our bound exhibits a poly-logarithmic dependence on time.  Among other things, this regret bound implies that CTRACE is {\it efficient} in the sense that the $(\epsilon,\delta)$-convergence time is bounded by a polynomial function of $1/\epsilon$ and $\log(1/\delta)$ with high probability.  We define the $(\epsilon,\delta)$-convergence time to be the first time when an estimate and all the future estimates following it are within an $\epsilon$-neighborhood of a true value with probability at least $1-\delta$.  Let us provide here some intuition for why CTRACE outperforms  CE.  First, regularization enhances exploration in a critical manner.  Without regularization, we are more likely to obtain overestimates of price impact.  Such an outcome abates trading and thus exploration, making it difficult to escape from the predicament.  Regularization reduces the chances of obtaining overestimates, and further, tends to yield underestimates that encourage active exploration.  Second, requiring a high degree of confidence reduces the chances of occasionally producing erratic estimates, which regularly arise with application of CE.  Such estimates can result in undesirable trades and/or reductions in the degree of exploration.  
 
It is also worth comparing CTRACE to a reinforcement learning algorithm recently proposed in \cite{Abbasi-YadkoriSzepesvari10} which appears well-suited for our problem.  This algorithm was designed to explore efficiently in a broader class of linear-quadratic control problems, and is based on the {\it principle of optimism in the face of uncertainty}.  \cite{Abbasi-YadkoriSzepesvari10} establish an $O ( \sqrt{T \log(1/\delta)} )$ regret bound that holds with probability at least $1-\delta$, 
where $T$ denotes time and some logarithmic terms are hidden.  
Our bound for CTRACE is on expected regret and exhibits a dependence on $T$ of $O(\log^2 T)$.
We also demonstrate via Monte Carlo simulation that CTRACE dramatically outperforms this algorithm. 

To summarize, the primary contributions of this paper include:
\begin{enumerate}[(a)]
\setlength{\itemsep}{0pt}
\item We propose a new method for simultaneous execution and learning -- the confidence-triggered regularized adaptive certainty equivalent (CTRACE) policy.
\item  We establish a finite-time expected regret bound for CTRACE that exhibits a poly-logarithmic dependence on time.  This bound implies that CTRACE is {\it efficient} in the sense that, with probability $1-\delta$, the $(\epsilon,\delta)$-convergence time is bounded by a polynomial function of $1/\epsilon$ and $\log(1/\delta)$.  
\item We demonstrate via Monte Carlo simulation that CTRACE outperforms the certainty equivalent policy and a reinforcement learning algorithm recently proposed by \cite{Abbasi-YadkoriSzepesvari10} which is designed to explore efficiently in linear-quadratic control problems.
\end{enumerate}

The organization of the rest of this paper is as follows: Section \ref{sec:problemFormulation} presents our problem formulation, establishes existence and uniqueness of an optimal solution to our problem, and defines performance measures that can be used to evaluate policies. In Section \ref{sec:CTRACE}, we propose CTRACE and derive a finite-time expected regret bound for CTRACE along with two properties: inter-temporal consistency and efficiency. Section \ref{sec:computationalAnalysis} is devoted to Monte Carlo simulation in which the performance of CTRACE is compared to that of two benchmark policies. Finally, we conclude this paper in Section \ref{sec:conclusion}. All proofs are provided in Appendix. Detailed proofs are available upon request.

\section{Problem Formulation} \label{sec:problemFormulation}

\subsection{Model Description}

\textbf{\textsf{\indent Decision Variable and Security Position }} We consider a trader who trades a single security over an infinite time horizon. She submits a market buy or sell order at the beginning of each period of equal length. $u_t \in \mathbb{R}$ represents the number of shares of the security to buy or sell at period $t$ and a positive (negative) value of $u_t$ denotes a buy (sell) order. Let $x_{t-1} \in \mathbb{R}$ denote the trader's pre-trade security position before placing an order $u_t$ at period $t$. Therefore, $x_t = x_{t-1} + u_t, \,\, t \geq 1$. 

\textbf{\textsf{Price Dynamics }} The absolute return of the security is given by
\begin{align} 
\Delta p_{t} &= p_{t} - p_{t-1} = g^\top f_{t-1} + \lambda^* u_{t} + \sum_{m=1}^{M} \gamma_m^* (d_{m,t} - d_{m,t-1}) + \epsilon_{t} \nonumber \\
d_{m,t} &\defeq \sum_{i=1}^{t} r_m^{t-i} u_i = r_m d_{m,t-1} + u_t, \quad d_t \defeq [  d_{1,t} \,\,\, \cdots \,\,\, d_{M,t} ]^\top. \label{eqn:transientImpact}
\end{align}
\noindent We will explain each term in detail as we progress. This can be viewed as a first-order Taylor expansion of a geometric model
\[
\log \left( \frac{ p_{t} }{ p_{t-1} } \right) = \tilde{g}^\top f_{t-1} + \tilde{\lambda}^* u_{t} + \sum_{m=1}^{M} \tilde{\gamma}_m^* (d_{m,t} - d_{m,t-1}) + \tilde{\epsilon}_{t}
\]
over a certain period of time, say, a few weeks in calendar time, which makes this approximation reasonably accurate for practical purposes. Although it is unrealistic that the security price can be negative with positive probability, our model nevertheless serves its practical purpose for the following reasons: Our numerical experiments conducted in Section \ref{sec:computationalAnalysis} show that price changes after a few weeks from now have ignorable impacts on a current optimal action. In other words, optimal actions for our infinite-horizon control problem appear to be quite close to those for a finite-horizon counterpart on a few week time scale. Furthermore, it turns out that in simulation we could learn a unknown price impact model fast enough to take actions that are close to optimal actions within a few weeks. Thus, learning based on our price dynamics model could also be justified. We will give concrete numerical examples later to support these notions.

\textbf{\textsf{Price Impact }} The term $\lambda^* u_t$ represents ``permanent price impact'' on the security price of a current trade. The permanent price impact is endogenously derived in \cite{Kyle85} from informational asymmetry between an informed trader and uninformed competitive market makers, and in \cite{Rosu09} from equilibrium of a limit order market where fully strategic liquidity traders dynamically choose limit and market orders.  \cite{HubermanStanzl04} prove that the linearity of a time-independent permanent price impact function is a necessary and sufficient condition for the absence of ``price manipulation'' and ``quasi-aribtrage'' under some regularity conditions. 

The term $\sum_{m=1}^{M} \gamma_m^* d_{m,t}$ indicates ``transient price impact'' that models other traders' responses to non-informative orders. For example, suppose that a large market buy order has arrived and other traders monitoring the market somehow realize that there is no definitive evidence for abrupt change in the fundamental value of the security. Then, they naturally infer that the large buy order came merely for some liquidity reason, and gradually ``correct'' the perturbed price into what they believe it is supposed to be by submitting counteracting selling orders. The dynamics of $d_{m,t}$ in (\ref{eqn:transientImpact}) indicates that the impact of a current trade on the security price decays exponentially over time, which is considered in \cite{ObizhaevaWang05} that incorporate the dynamics of supply and demand in a limit order market to optimal execution strategies. In \cite{Gatheral10}, it is shown that the exponentially decaying transient price impact is compatible only with a linear instantaneous price impact function in the absence of ``dynamic arbitrage.'' 

\textbf{\textsf{Observable Return-Predictive Factors }} We assume that there are multiple observable return-predictive factors that affect the absolute return of the security as in \cite{GarleanuPedersen09}. Those factors could be macroeconomic factors such as gross domestic products (GDP), inflation rates and unemployment rates, 
security-specific factors such as P/B ratio, P/E ratio and lagged returns, 
or prices of other securities that are correlated with the security price. In our price dynamics model, $f_t \in \mathbb{R}^{K}$ denotes these factors and $g \in \mathbb{R}^{K}$ denotes factor loadings. The term $g^\top f_{t-1}$ represents predictable excess return or ``alpha.'' We assume that $f_t$ is a first-order vector autoregressive process $f_{t} = \Phi f_{t-1} + \omega_{t}$ where $\Phi \in \R^{K \times K}$ is a stable matrix that has all eigenvalues inside a unit disk and $\omega_t \in \R^{K}$ is a martingale difference sequence adapted to the filtration $\{ \mathcal{F}_t = \sigma(\{ x_0, d_0, f_0, \omega_1, \ldots, \omega_t, \epsilon_1, \ldots, \epsilon_t \}) \}$. We further assume that $\omega_t$ is bounded almost surely, $\text{i.e.}$ $\| \omega_t \| \leq C_\omega \,\, a.s. $ for all $t \geq 1$ for some deterministic constant $C_\omega$, and $\text{Cov}[\omega_t | \mathcal{F}_{t-1}] = \Omega \in \R^{K \times K}$ being positive definite and independent of $t$. 

\textbf{\textsf{Unpredictable Noise }} The term $\epsilon_t$ represents random fluctuations that cannot be accounted for by price impact and  observable return-predictive factors. We assume that $\epsilon_t$ is a martingale difference sequence adapted to the filtration $\{ \mathcal{F}_t \}$, and independent of $x_0$, $d_0$, $f_0$ and $\omega_\tau$ for any $\tau \geq 1$. Also, $\E[ \epsilon_t^2 | \mathcal{F}_{t-1}] = \Sigma_\epsilon \in \R$ being independent of $t$. Finally, each $\epsilon_t$ is assumed to be sub-Gaussian, $\text{i.e.}$, $\E[ \textrm{exp}(a \epsilon_{t}) | \mathcal{F}_{t-1}] \leq \textrm{exp}(C_\epsilon^2 a^2 /2), \,\, \forall t \geq 1, \,\, \forall a \in \mathbb{R}$ for some $C_\epsilon > 0$.   

\textbf{\textsf{Policy }} A policy is defined as a sequence $\pi = \{ \pi_1, \pi_2, \ldots \}$ of functions where $\pi_t$ maps the trader's information set at the beginning of period $t$ into an action $u_t$. The trader observes $f_{t-1}$ and $p_{t-1}$ at the end of period $t-1$ and thus her information set at the beginning of period $t$ is given by $\mathcal{I}_{t-1} = \{ x_0, d_0, f_0, \ldots, f_{t-1}, p_0, \ldots, p_{t-1} \}$. A policy $\pi$ is {\it admissible} if $z_t \defeq [x_t \,\,\, d_{t}^\top \,\,\, f_t^\top]^\top$ generated by $u_t = \pi_t(\mathcal{I}_{t-1})$ satisfies $\lim_{T \rightarrow \infty} \| z_T \|^2/T = 0$. A set of admissible policies is denoted by $\Pi$.

\textbf{\textsf{Objective Function }} The trader's objective is to maximize expected average ``risk-adjusted'' profit defined as
\[ 
\liminf_{T \rightarrow \infty} \,\, \E \left[ \frac{1}{T}  \sum_{t=1}^{T} \left( \Delta p_{t} x_{t-1} - \rho \Sigma_\epsilon x_t^2 \right) \right] 
\]
where the first term $\Delta p_{t} x_{t-1}$ indicates change in book value and the second term $\rho \Sigma_\epsilon x_{t}^2$ a quadratic penalty for her non-zero security position in the next period that reflects her risk aversion. $\rho$ is a risk-aversion coefficient that quantifies the extent to which the trader is risk-averse. 

\textbf{\textsf{Assumptions }} The following is a list of assumptions on which our analysis is based throughout this paper. Let $\theta^{*} \defeq \left[ \lambda^{*} \,\, \gamma^{*}_{1} \,\, \ldots \,\, \gamma^{*}_{M} \right]^\top \in \mathbb{R}^{M+1}$. We will make two more assumptions as we progress. 
\begin{assumption} \label{assum:one}
\begin{enumerate}[(a)]
\setlength{\itemsep}{0pt}
\item The price impact coefficients $\theta^{*}$ are unknown to the trader. Note that they can be learned only through executed trades.
\item The factor loadings $g$ are known to the trader. This is a reasonable assumption since they can be learned by observing prices without any transaction. 
\item The decaying rates $r \defeq [r_1, \,\, \ldots, \,\, r_M]^\top \in [0, 1)^M$ of the transient price impact are known to the trader and all the elements are distinct. In practice, they are definitely not known a priori. However, it can be handled effectively for practical purposes by using a sufficiently dense $r$ with a large $M$ so that potential bias induced by modeling mismatch can be greatly reduced at the expense of increased variance, which can be reduced by regularization.
\item $\theta^{*} \in \Theta \defeq \{ \theta \in \R^{M+1} : 0 \leq \theta \leq \theta_\text{max}, \,\, \1^\top \theta \geq \beta \}$ for some $\theta_\text{max} > 0$ component-wise and some $\beta > 0$. The constraint $\1^\top \theta \geq \beta$ is imposed to capture non-zero execution costs in practice. Note that $\Theta$ is compact and convex.
\end{enumerate}
\end{assumption}

\textbf{\textsf{Notations }} $\| \cdot \|$ and $\| \cdot \|_F$ denote the $\ell_2$-norm and the Frobenius norm of a matrix, respectively. $a \vee b$ and $a \wedge b$ denote $\max\{ a, b\}$ and $\min\{ a, b\}$, respectively. For a symmetric matrix $A$, $A \succ 0$ means that $A$ is positive definite and $A \succeq 0$ means that $A$ is positive semidefinite. $\lambda_{\text{min}}(A)$ indicates the smallest eigenvalue of $A$. $(A)_{ij}$ of a matrix $A$ indicates the entry of $A$ in the $i$th row and in the $j$th column. $(v)_i$ of a vector $v$ indicates the $i$th entry of $v$. $\text{diag}(v)$ of a vector $v$ denotes a diagonal matrix whose $i$th diagonal entry is $(v)_i$. $A_{*,j}$ denotes the $j$th column of $A$ and $A_{i:j,k}$ indicates a segment of the $k$th column of $A$ from the $i$th entry to the $j$th entry. $\1\{ \mathcal{B} \}$ denotes an indicator function on the event $\mathcal{B}$.

\subsection{Existence of Optimal Solution} \label{sec:existenceOptSoln}

Now, we will show that there exists an optimal policy among admissible policies that maximizes expected average risk-adjusted profit. For convenience, we will consider the following minimization problem that is equivalent to maximize expected average risk-adjusted profit.
\begin{align*} 
\min_{ \pi \in \Pi} \,\, \limsup_{T \rightarrow \infty} \,\, \E \left[ \frac{1}{T}  \sum_{t=1}^{T} \left(\rho \Sigma_\epsilon x_{t}^2 -  \Delta p_{t} x_{t-1} \right) \right]
\end{align*}
We call the negative of average risk-adjusted profit ``average cost.'' This problem can be expressed as a discrete-time linear quadratic control problem
\[
\min_{ \pi \in \Pi } \,\, \limsup_{T \rightarrow \infty} \,\, \E \left[ \frac{1}{T}  \sum_{t=1}^{T} 
\left[ \begin{array}{cc} z_{t-1}^\top & u_t \end{array} \right]
\left[ \begin{array}{cc} Q & S \\ S^\top & R \end{array} \right]
\left[ \begin{array}{c} z_{t-1} \\ u_t \end{array} \right] \right] \,\, \text{s.t.} \,\, z_t = A z_{t-1} + B u_t + W_t, \,\, u_t = \pi_t (\mathcal{I}_{t-1})
\]
where $z_t = [ x_t \,\,\, d_t^\top \,\,\, f_t^\top ]^\top$, $v = [ 0 \,\,\, \gamma^{*\top} (\text{diag}(r) - I) \,\,\, g^\top ]^\top$, $\gamma^{*} = [ \gamma_1^{*} \,\,\, \cdots \,\,\, \gamma_M^{*} ]^\top$, $e_1 = [ 1 \,\,\, 0 \,\,\, \cdots \,\,\, 0 ]^\top$,
\[
Q = \rho \Sigma_\epsilon e_1 e_1^\top - \frac{1}{2} (v e_1^\top + e_1 v^\top), \quad S = \rho \Sigma_\epsilon e_1 - \frac{1}{2} (\lambda^* + \gamma^{*\top} \1) e_1, \quad R = \rho \Sigma_\epsilon, 
\]
\[
A = 
\left[ \begin{array}{ccc} 
1 & 0 & 0 \\
0 & \text{diag}(r) & 0 \\
0 & 0 & \Phi 
\end{array} \right], \quad 
B = \left[ \begin{array}{c} 1 \\ \1 \\ 0 \end{array} \right], \quad
W_t = \left[ \begin{array}{c} 0 \\ 0 \\ \omega_{t} \end{array} \right], \quad
\tilde{\Omega} \defeq \text{Cov}[W_t] = \left[ \begin{array}{ccc} 0 & 0 & 0 \\ 0 & 0 & 0 \\ 0 & 0 & \Omega \end{array} \right].
\]
Note that $R$ is strictly positive but $Q$ is not necessarily positive semidefinite. Therefore, special care should be taken in order to prove the existence of an optimal policy. We start with a well-known Bellman equation for average-cost linear quadratic control problems 
\begin{equation} \label{eqn:BellmanEqn}
H(z_{t-1}) + h = \min_{u_t}\,\, \E \left[ \rho \Sigma_\epsilon (x_{t-1} + u_{t})^2 - \Delta p_{t} x_{t-1} + H(z_{t}) \right]
\end{equation}
where $H(\cdot)$ denotes a differential value function and $h$ denotes minimum average cost. It is natural to conjecture $H(z_{t}) = z_{t}^\top P z_{t}$. Plugging it into (\ref{eqn:BellmanEqn}), we can obtain a discrete-time Riccati algebraic equation
\begin{equation} \label{eqn:dare}
P = A^\top P A + Q - (S^\top + B^\top P A)^\top (R + B^\top P B)^{-1} (S^\top + B^\top P A)
\end{equation}
with a second-order optimality condition $R + B^\top P B > 0$. The following theorem characterizes an optimal policy among admissible policies that minimizes expected average cost, and proves existence and uniqueness of such an optimal policy.
\begin{theorem} \label{thm:optPolicyExistenceUniqueness}
For any $\theta^* \in \Theta$, there exists a unique symmetric solution $P$ to (\ref{eqn:dare}) that satisfies $R + B^\top P B > 0$ and $\rho_{\text{sr}} (A + B L)  < 1$ where $L = -(R + B^\top P B)^{-1} (S^\top + B^\top P A)$ and $\rho_{\text{sr}} (\cdot)$ denotes a spectral radius. Moreover, a policy $\pi = (\pi_1, \pi_2, \ldots )$ with $\pi_t(\mathcal{I}_{t-1}) = L z_{t-1}$ is an optimal policy among admissible policies that attains minimum expected average cost $\text{tr}(P \tilde{\Omega})$.
\end{theorem}

For ease of exposition, we define some notations: $P(\theta)$ denotes a unique symmetric stabilizing solution to (\ref{eqn:dare}) with $\theta^* = \theta$. $L(\theta) \defeq  -(R + B^\top P(\theta) B)^{-1} (S(\theta)^\top + B^\top P(\theta) A)$ denotes a gain matrix for an optimal policy with $\theta^* = \theta$, $G(\theta) \defeq A + B L(\theta)$ denotes a closed-loop system matrix with $\theta^* = \theta$, and $U(\theta) \defeq \1 L(\theta) + \left[A-I \,\,\, O \right]$ denotes a linear mapping from $z_{t-1}$ to a regressor $\psi_t$ used in least-squares regression for learning price impact, $\text{i.e.}$ $\psi_t = U(\theta) z_{t-1}$. Having these notations, we make two assumptions about $L(\theta)$ as follows. Indeed, we can verify through closed-form solutions that these assumptions hold in a special case which will be discussed in Subsection \ref{subsec:closedForm}.

\begin{assumption} \label{assum:optimalPolicy}
\begin{enumerate}[(a)]
\setlength{\itemsep}{0pt}
\item There exists $C_L > 0$ such that $\| L(\theta_1) - L(\theta_2) \| \leq C_L \| \theta_1 - \theta_2 \| $ for any $\theta_1, \theta_2 \in \Theta$. 
\item $(L(\theta))_1 \neq 0$ and $(L(\theta))_{M+2} \neq 0$ for any $\theta \in \Theta$
\end{enumerate}
\end{assumption}

Using Assumption \ref{assum:optimalPolicy}, we can obatin an upper bound on $\| z_t \|$ uniformly over $\theta \in \Theta$ and $t \geq 0$.
\begin{lemma} \label{lem:uniformBound}
For any $0 < \xi < 1$, there exists $N \in \mathbb{N}$ being independent of $\theta$ such that $\| G^N(\theta) \| \leq \xi$ for all $\theta \in \Theta$. Thus, $\max_{0 \leq i \leq N-1} \sup_{\theta \in \Theta} \| G^i(\theta) \| \defeq C_g$ is finite. For any fixed $\theta \in \Theta$, $\| z_t \| \leq C_g \| z_0 \| + C_g C_\omega / (\xi (1 - \xi^{1/N})) \defeq C_z, \,\, \forall t \geq 0 \,\, a.s.$ where $z_t = G(\theta) z_{t-1} + W_t$. Moreover, $\sup_{\theta \in \Theta} \| U(\theta) \| \leq C_g + 1$. 
\end{lemma}

Note that Lemma \ref{lem:uniformBound} can be applied only when $\theta$ is fixed over time. From now on, we assume $\| z_0 \| \leq 2 C_g C_\omega / (\xi (1-\xi^{1/N}))$ without loss of generality otherwise we can always set $C_g$ to be greater than $\| z_0 \| \xi (1-\xi^{1/N}) / (2 C_\omega)$. 

\begin{figure}[t]
\hspace{-1.1cm}
\includegraphics[scale=0.55]{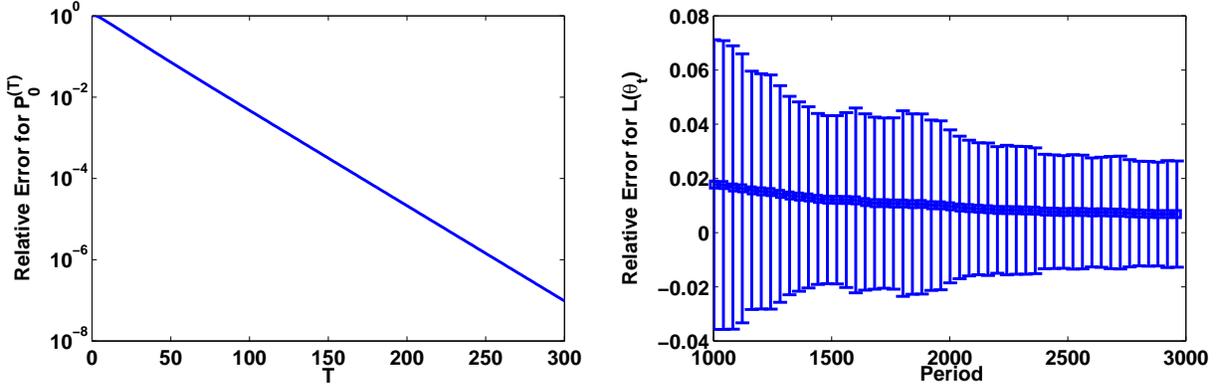}
\caption{(Left) Relative error for $P_T$: $T = 300$ corresponds to 3.8 trading days. (Right) Relative error for $L(\theta_t)$ from CTRACE: Period 3000 corresponds to 38 trading days. The verical bars represent two standard errors. In both figures, the simulation setting in Section \ref{sec:computationalAnalysis} is used. }
\label{fig:mixingLearningTime}
\end{figure}

Finally, we present concrete numerical examples that support the validity of our price model as an approximation of the geometric model for practical purposes. As we discussed earlier, our numerical experiments conducted in Section \ref{sec:computationalAnalysis} show that our infinite-horizon control problem could be approximated accurately by a finite-time control problem with a time horizon on a few week time scale. To be more precise, we define {\it relative error for $P_0^{(T)}$} as $\| P_{0}^{(T)} - P \| / \| P \|$ where $P_t^{(T)}$ denotes a coefficient matrix of a quadratic value function at period $t$ for a finite-horizon control problem with a terminal period $T$, and $P$ denotes a coefficient matrix of a quadratic value function for our infinite-horizon control problem. As shown in Figure \ref{fig:mixingLearningTime}, the relative error for $P_0^{(T)}$ appears to decrease exponentially in $T$ and the relative error for $P_0^{(300)}$ is almost $10^{-7}$ where $T = 300$ corresponds to 3.8 trading days. 

Furthermore, we could learn unknown $\theta^*$ fast enough to take actions that are close to optimal actions on a required time scale. An action from a current estimate could be quite close to an optimal action even if estimation error for the current estimate is large, especially in cases where a few ``principal components'' of $L(\theta)$ with large directional derivatives with respect to $\theta$ are learned accurately. To be more precise, we define {\it relative error for $L(\theta_t)$} as 
\[
\frac{\text{E}[(L(\theta_t) z_{t-1}^* - L(\theta^*) z_{t-1}^* )^2]}{\text{E}[(L(\theta^*) z_{t-1}^*)^2]} = \frac{(L(\theta_t) - L(\theta^*)) \Pi_{zz}(\theta^*) (L(\theta_t) - L(\theta^*))^\top}{L(\theta^*) \Pi_{zz}(\theta^*) L(\theta^*)^\top}
\]
where $z_t^*$ is a stationary process generated by $u_t^* = L(\theta^*) z_{t-1}^*$ and $\Pi_{zz}(\theta^*) = E[z_t^* z_t^{* \top}]$. The relative error for $L(\theta_t)$ indicates how different an action from an estimate $\theta_t$ is than an optimal action from the true value $\theta^*$. Figure \ref{fig:mixingLearningTime} shows how the relative error for $L(\theta_t)$ evolves over time with two-standard-error bars when $\theta_t$'s are obtained from a new policy that we will propose in Section \ref{sec:CTRACE}. As you can see, all the approximate 95\%-confidence intervals lie within $\pm 3\%$ range after Period 2500 that corresponds to 32 trading days. It implies that actions from estimates learned over a few weeks could be sufficiently close to optimal actions.

\subsection{Closed-Form Solution: A Single Factor and Permanent Impact Only} \label{subsec:closedForm}

When we consider only the permanent price impact and a single observable factor, we can derive an exact closed-form $P$ and $L$ as follows.
\[
P_{xx} = \frac{\lambda^* - \rho \Sigma_\epsilon + \sqrt{ 2 \lambda^* \rho \Sigma_\epsilon + (\rho \Sigma_\epsilon)^2}}{2}
\]
\[
P_{xf} = \frac{-g \lambda^*}{(1-\Phi) \lambda^* - \Phi \rho \Sigma_\epsilon + \Phi \sqrt{2 \lambda^* \rho \Sigma_\epsilon + (\rho \Sigma_\epsilon)^2} }
\]
\[
P_{ff} = \frac{- g^2 \Phi^2}{2 (1-\Phi^2) \left( (1-\Phi)^2 \lambda^* + (1+\Phi^2) \rho \Sigma_\epsilon + (1-\Phi^2) \sqrt{2 \lambda^* \rho \Sigma_\epsilon + (\rho \Sigma_\epsilon)^2} \right)}
\]
\[
L_x = \frac{-2 \rho \Sigma_\epsilon}{\rho \Sigma_\epsilon + \sqrt{2 \lambda^* \rho \Sigma_\epsilon + (\rho \Sigma_\epsilon)^2}}
\]
\[
L_f = \frac{g \Phi }{(1-\Phi) \lambda^* + \rho \Sigma_\epsilon + \sqrt{2 \lambda^* \rho \Sigma_\epsilon + (\rho \Sigma_\epsilon)^2}}
\]
Although this is a special case of our general setting, we can get useful insights into the effect of permanent price impact coefficient $\lambda^*$ on various quantities. Here are some examples:
\begin{itemize}
\setlength{\itemsep}{0pt}
\item $| L_x |$ and $| L_f |$ are strictly decreasing in $\lambda^*$. 
\item $\lim_{\lambda^* \rightarrow 0} L_x = -1$, $\lim_{\lambda^* \rightarrow \infty} L_x = 0$. 
\item $\lim_{\lambda^* \rightarrow 0} L_f = g \Phi / (2 \rho \Sigma_\epsilon)$, $\lim_{\lambda^* \rightarrow \infty} L_f = 0$.
\item The expected average risk-adjusted profit $-P_{ff} \Omega$ is strictly decreasing in $\lambda^*$.
\item $\lim_{\lambda^* \rightarrow 0} (-P_{ff} \Omega) = g^2 \Phi^2 \Omega / ( 4(1-\Phi^2) \rho \Sigma_\epsilon)$, $\lim_{\lambda^* \rightarrow \infty} (-P_{ff} \Omega) = 0$.
\end{itemize}

\subsection{Performance Measure: Regret} \label{subsec:regret}

In this subsection, we define a performance measure that can be used to evaluate policies. For notational simplicity, let $L^{*} = L(\theta^{*})$, $G^{*} = G(\theta^{*})$ and $P^* = P(\theta^{*})$. Using (\ref{eqn:dare}), we can show that 
\vspace{0in}
\[
\begin{split}
&J_T^\pi (z_0 | \mathcal{F}_{T} ) \defeq \sum_{t=1}^{T} \left\{ \rho \Sigma_\epsilon (x_{t-1} + \pi_{t}(\mathcal{I}_{t-1}) )^2 - \Delta p_t x_{t-1} \right\} \\
& \quad = z_0^\top P^* z_0 - z_T^\top P^* z_T + 2 \sum_{t=1}^{T} (A z_{t-1} + B \pi_{t}(\mathcal{I}_{t-1}) )^\top P^* W_t + \sum_{t=1}^{T} W_t^\top P^* W_t - \sum_{t=1}^{T} x_{t-1} \epsilon_t \\
& \qquad + \sum_{t=1}^{T} (\pi_{t}(\mathcal{I}_t) - L^{*} z_{t-1})^\top (R + B^\top P^* B) (\pi_{t}(\mathcal{I}_{t-1}) - L^{*} z_{t-1}) \quad \text{for any policy } \pi.
\end{split}
\]
First, we define {\it pathwise regret} $R_T^\pi (z_0 | \mathcal{F}_{T} )$ of a policy $\pi$ at period $T$ as
$J_T^\pi (z_0 | \mathcal{F}_{T} ) - J_T^{\pi^{*}} (z_0 | \mathcal{F}_{T} )$ where $\pi_t^{*}(\mathcal{I}_{t-1}) = L^{*} z_{t-1}^{*}$ and $z_t^{*} = G^* z_{t-1}^{*} + W_t$ with $z_0^{*} = z_0$.
In other words, the pathwise regret of a policy $\pi$ at period $T$ amounts to excess costs accumulated over $T$ periods when applying $\pi$ relative to when applying the optimal policy $\pi^{*}$.
By definition of $\pi^{*}$, the pathwise regret of a policy $\pi$ at period $T$ can be expressed as
\[
\begin{split} 
&R_T^\pi (z_0 | \mathcal{F}_{T} ) = z_T^{* \top} P^* z_T^{*}  - z_T^\top P^* z_T + \sum_{t=1}^{T} (\pi_t(\mathcal{I}_{t-1}) - L^{*} z_{t-1})^\top (R + B^\top P^* B) (\pi_t(\mathcal{I}_{t-1})  - L^{*} z_{t-1}) \\
& \quad + 2 \sum_{t=1}^{T} ((A z_{t-1} + B \pi_t(\mathcal{I}_{t-1}) ) - (A + B L^{*}) z_{t-1}^{*}  )^\top P^* W_t + \sum_{t=1}^{T} (x_{t-1}^{*} - x_{t-1}) \epsilon_t.
\end{split}
\]
Second, we define {\it expected regret} $\bar{R}_T^{\pi} (z_0)$ of a policy $\pi$ at period $T$ as
$\E[ R_T^\pi(z_0 | \mathcal{F}_{T} ) ]$. Taking expectation of pathwise regret, we can obtain a more concise expression for expected regret because the last two terms vanish by the law of total expectation. Hence, we have
\[
\bar{R}_T^{\pi} (z_0) = \E [ z_T^{* \top} P^* z_T^{*}  - z_T^\top P^* z_T ] + \E \left[ \sum_{t=1}^{T} (\pi_t(\mathcal{I}_{t-1}) - L^{*} z_{t-1})^\top (R + B^\top P^* B) (\pi_t(\mathcal{I}_{t-1}) - L^{*} z_{t-1}) \right].
\]
Finally, we define {\it relative regret} $\tilde{R}_T^{\pi} (z_0)$ of a policy $\pi$ at period $T$ as
$\bar{R}_T^{\pi} (z_0) / | \textrm{tr}(P^{*} \tilde{\Omega}) |$ where $\textrm{tr}(P^{*} \tilde{\Omega})$ is minimum expected average cost for $\theta^{*}$. Our choice of performance measure will be either expected regret or relative regret in the rest of this paper.

\section{\fontsize{13.5}{13.5}\selectfont Confidence-Triggered Regularized Adaptive Certainty Equivalent Policy} \label{sec:CTRACE}

Our problem can be viewed as a special case of reinforcement learning, which focuses on sequential decision-making problems in which unknown properties of an environment must be learned in the course of taking actions. It is often emphasized in reinforcement learning that longer-term performance can be greatly improved by making decisions that explore the environment efficiently at the expense of suboptimal short-term behavior. In our problem, a price impact model is unknown, and submission of large orders can be considered exploratory actions that facilitate learning. 

Certainty equivalent control (CE) represents one extreme where at any time, current point estimates are assumed to be correct and actions are made accordingly. Although learning is carried out with observations made as the system evolves, no decisions are designed to enhance learning. Thus, this is an instance of pure exploitation of current knowledge. In our problem, CE estimates the unknown price impact coefficients $\theta^*$ at each period via least-squares regression using available data, and makes decisions that maximize expected average risk-adjusted profit under an assumption that the estimated model is correct. That is, an action $u_t$ for CE is given by $u_t = L(\tilde{\theta}_{t-1}) z_{t-1}$ where $\tilde{\theta}_{t-1} = \argmin_{\theta \in \Theta} \sum_{i=1}^{t-1} \left( (\Delta p_i - g^\top f_{i-1}) - \psi_i^\top \theta  \right)^2$ with a regressor $\psi_i = [ u_i \,\,\, (d_i - d_{i-1})^\top ]^\top$. 

An important question is how aggressively the trader should explore to learn $\theta^*$. Unlike many other reinforcement learning problems, a fairly large amount of exploration is naturally induced by exploitative decisions in our problem. That is, regular trading activity triggered by the return-predictive factors $f_t$ excites the market regardless of whether or not she aims to learn price impact. Given sufficiently large factor variability, the induced exploration might adequately resolve uncertainties about price impact. However, we will demonstrate by proposing a new exploratory policy that executing trades to explore beyond what would naturally occur through the factor-driven exploitation can result in significant benefit.  

Now, let us formally state that exploitative actions triggered by the return-predictive factors induce a large degree of exploration that could yield strong consistency of least-squares estimates. It is worth noting that pure exploitation is not sufficient for strong consistency in other problems such as \cite{LaiWei86} and \cite{ChenGuo86}.

\begin{lemma} \label{lem:persistentExcitationFixed}
For any $\theta \in \Theta$, let $u_t = L(\theta) z_{t-1}$, $z_t = G(\theta) z_{t-1} + W_t$ and $\psi_t^\top = \left[ u_t \,\,\, (d_t - d_{t-1})^\top \right] = (U(\theta) z_{t-1})^\top$. Also, let $\Pi_{zz}(\theta)$ denote a unique solution to $\Pi_{zz}(\theta) = G(\theta) \Pi_{zz}(\theta) G(\theta)^\top +  \tilde{\Omega}$. Then, 
\begin{align} \label{eqn:ergodic}
\lim_{T \rightarrow \infty} \frac{1}{T} \sum_{t=1}^{T} \psi_t \psi_t^\top = U(\theta) \Pi_{zz}(\theta) U(\theta)^\top \succ 0 \,\,\, a.s. 
\end{align}
\end{lemma}

Moreover, we can show that $\Pi_{zz}(\theta)$ is continuous on $\Theta$ by proving uniform convergence of $\E \left[ \frac{1}{T} \sum_{t=1}^{T} z_{t-1} z_{t-1}^\top \right]$ to $\Pi_{zz}(\theta)$ on $\Theta$. Continuity leads to $\underline{\lambda}_{\psi \psi}^{*} \defeq \inf_{\theta \in \Theta} \lambda_{\text{min}} \left( U(\theta) \Pi_{zz}(\theta) U(\theta)^\top\right) > 0$ which will be used later. 

\begin{corollary} \label{cor:extremums}
$\Pi_{zz}(\theta)$ is continuous on $\Theta$ and $\underline{\lambda}_{\psi \psi}^{*} \defeq \inf_{\theta \in \Theta} \lambda_{\text{min}} \left( U(\theta) \Pi_{zz}(\theta) U(\theta)^\top\right) > 0$. 
\end{corollary} 

Lemma \ref{lem:persistentExcitationFixed} implies that $\lambda_{\text{min}} \left( \sum_{t=1}^{T} \psi_{t} \psi_{t}^\top \right)$ increases linearly in time $T$ $\text{a.s.}$ asymptotically. In addition, we can obtain a similar result for a finite-sample case: There exists a finite, deterministic constant $T_1(\theta,\delta)$ such that $\lambda_{\text{min}} \left( \sum_{t=1}^{T} \psi_{t} \psi_{t}^\top \right)$ grows linearly in time $T$ for all $T \geq T_1(\theta,\delta)$ with probability at least $1-\delta$. This is a crucial result that will be used for bounding above ``$(\epsilon,\delta)$-convergence time'' later. It is formally stated in the following lemma.

\begin{lemma} \label{lem:linearGrowthFixed}
For any $\theta \in \Theta$, let $u_t = L(\theta) z_{t-1}$, $z_t = G(\theta) z_{t-1} + W_t$ and $\psi_t^\top = \left[ u_t \,\,\, (d_t - d_{t-1})^\top \right] = (U(\theta) z_{t-1})^\top$. Then, there exists an event $\mathcal{B}(\delta)$ such that on $\mathcal{B}(\delta)$ with $\text{Pr}(\mathcal{B}(\delta)) \geq 1-\delta$
\[
\frac{7}{8} U(\theta) \Pi_{zz}(\theta) U(\theta)^\top \preceq \frac{1}{T} \sum_{t=1}^{T} \psi_{t} \psi_{t}^\top \preceq \frac{17}{16} U(\theta) \Pi_{zz}(\theta) U(\theta)^\top \quad \forall T \geq T_1(\theta,\delta) \quad \text{where}
\]
{\small
\[
T_1(\theta,\delta) = 4 \left( \frac{32 (C_z C_g)^2 (M+K+1)}{\xi^2 (1-\xi^{\frac{2}{N}}) \lambda_{\text{min}}(\Pi_{zz}(\theta)) } \right)^2 \log \left( \frac{(M+K+2)^4}{432 \delta^2} \right) \vee 8 \left( \frac{32 (C_z C_g)^2 (M+K+1)}{\xi^2 (1-\xi^{\frac{2}{N}}) \lambda_{\text{min}}(\Pi_{zz}(\theta)) } \right)^3 \vee 216. 
\]
}
\end{lemma}	

Furthermore, we can extend Lemma \ref{lem:persistentExcitationFixed} in such a way that $\lambda_{\text{min}} \left( \sum_{t=1}^{T} \psi_{t} \psi_{t}^\top \right)$ still increases to infinity linearly in time $T$ for time-varying $\{ \theta_t \}$ adapted to $\{ \sigma(\mathcal{I}_{t}) \}$ as long as $\theta_t$  remains sufficiently close to a fixed $\theta \in \Theta$ for all $t \geq 0$. Here, $\sigma(\mathcal{I}_{t})$ denotes a $\sigma$-algebra generated by $\mathcal{I}_{t}$ and $\theta_t$ is $\sigma(\mathcal{I}_{t})$-measurable for each $t$. 

\begin{lemma} \label{lem:persistentExcitationFloating}
Consider any $\theta \in \Theta$ and $\{ \theta_t \in \Theta \}$ adapted to $\{ \sigma(\mathcal{I}_t) \}$ such that $\| \theta_t - \theta \| \leq \frac{\eta}{\sqrt{M+1} C_L} \,\, a.s.$
\[
\text{where} \quad \eta = \left( \frac{\nu^3 (1-\nu^{\frac{1}{N}})^3 \lambda_{\text{min}}(\Pi_{zz}(\theta))}{42 N C_g^{N+1} C_\omega^2} \wedge \frac{\nu^3 (1-\nu^{\frac{1}{N}})^3 \lambda_{\text{min}}(U(\theta) \Pi_{zz}(\theta) U(\theta)^\top)}{42 N C_g^{N+1} C_\omega^2 (1 + \| U(\theta) \|)^2} \wedge \frac{\nu - \xi}{N C_g^{N-1}} \right)
\]
for all $t \geq 0$ and any $\nu \in (\xi, 1)$.
Let $u_t = L(\theta_{t-1}) z_{t-1}$, $z_t = G(\theta_{t-1}) z_{t-1} + W_t$ and $\psi_t^\top = \left[ u_t \,\,\, (d_t - d_{t-1})^\top \right] = (U(\theta_{t-1}) z_{t-1})^\top$. Then, 
\[
\liminf_{T \rightarrow \infty} \,\, \frac{1}{T} \sum_{t=1}^{T} \psi_t \psi_t^\top \succeq \frac{ \lambda_{\text{min}}( U(\theta) \Pi_{zz}(\theta) U(\theta)^\top)}{2} I \,\,\, a.s. 
\]
\end{lemma}

Similarly to Lemma \ref{lem:linearGrowthFixed}, we can obtain a finite-sample result for Lemma \ref{lem:persistentExcitationFloating}. This result will provide with a useful insight into how our new exploratory policy operates in the long term. 

\begin{lemma} \label{lem:linearGrowthFloating}
Consider $\{ \theta_t \in \Theta \}$ defined in Lemma \ref{lem:persistentExcitationFloating}. Let $u_t = L(\theta_{t-1}) z_{t-1}$, $z_t = G(\theta_{t-1}) z_{t-1} + W_t$ and $\psi_t^\top = \left[ u_t \,\,\, (d_t - d_{t-1})^\top \right] = (U(\theta_{t-1}) z_{t-1})^\top$. Then, for any $0 < \delta < 1$ on the event $\mathcal{B}(\delta)$ in Lemma \ref{lem:linearGrowthFixed} with $\text{Pr}(\mathcal{B}(\delta)) \geq 1 - \delta$
\[
\lambda_{\text{min}} \left( \frac{1}{T} \sum_{t=1}^{T} \psi_t \psi_t^\top  \right) \geq \frac{3}{8} \lambda_{\text{min}}(U(\theta) \Pi_{zz}(\theta) U(\theta)^\top), \quad \forall T \geq T_1(\theta, \delta) \vee \frac{3 \| z_0 \| (2 C_\omega + \| z_0 \|)}{C_\omega^2}.
\]
\end{lemma}	

\begin{algorithm}[b!]
\caption{CTRACE}
\label{alg:CTRACE}
\begin{algorithmic}[1]
\REQUIRE $\theta_0$, $x_0$, $d_0$, $r$, $g$, $\kappa$, $C_v$, $\tau$, $L(\cdot)$, $\theta_{\text{max}}$, $\{ p_t \}_{t=0}^{\infty}$, $\{ f_t \}_{t=0}^{\infty}$ 
\ENSURE $\{ u_t \}_{t=1}^{\infty}$ 
\STATE $V_0 \gets \kappa I$, $t_0 \gets 0$, $i \gets 1$
\FOR {$t = 1, 2, \ldots$}
	\STATE $u_t \gets L(\theta_{t-1}) z_{t-1}$, $x_t \gets x_{t-1} + u_t$, $d_t \gets \text{diag}(r) d_{t-1} + \1 u_t$
	\STATE $\psi_t \gets [ u_t \,\,\, (d_t - d_{t-1})^\top]^\top$, $V_t \gets V_{t-1} + \psi_t \psi_t^\top$
	\IF {$\lambda_{\text{min}}(V_{t}) \geq \kappa + C_v t$ and $t \geq t_{i-1} + \tau$ }
		\STATE $\theta_t \gets \argmin_{\theta \in \Theta} \sum_{i=1}^{t} \left( (\Delta p_i - g^\top f_{i-1}) - \psi_i^\top \theta  \right)^2 + \kappa \| \theta \|^2$, $t_i \gets t$, $i \gets i+1$
	\ELSE
		\STATE $\theta_t \gets \theta_{t-1}$
	\ENDIF		
\ENDFOR
\end{algorithmic}
\end{algorithm}

It is challenging to guarantee that all estimates generated by CE are sufficiently close to one another uniformly over time so that  Lemma \ref{lem:persistentExcitationFloating} and Lemma \ref{lem:linearGrowthFloating} can be applied to CE. In particular, CE is subject to overestimation of price impact that could be considerably detrimental to trading performance. The reason is that overestimated price impact discourages submission of large orders and thus it might take a while for the trader to realize that price impact is overestimated due to reduced ``signal-to-noise ratio.'' To address this issue, we propose the {\it confidence-triggered regularized adaptive certainty equivalent} policy (CTRACE) as presented in Algorithm \ref{alg:CTRACE}. CTRACE can be viewed as a generalization of CE and deviates from CE in two ways: (1) $\ell_2$ regularization is applied in least-squares regression, (2) coefficients are only updated when a certain measure
of confidence exceeds a pre-specified threshold and a minimum inter-update time has elapsed. Note that CTRACE reduces to CE as the regularization penalty $\kappa$ and the threshold $C_v$ tend to zero, and the minimum inter-update time $\tau$ tends to one. 

Regularization induces active exploration in our problem by penalizing the $\ell_2$-norm of price impact coefficients as well as reduces the variance of an estimator. Without regularization, we are more likely to obtain overestimates of price impact. Such an outcome attenuates trading intensity and thereby makes it difficult to escape from the misjudged perspective on price impact. Regularization decreases the chances of obtaining overestimates by reducing the variance of an estimator and furthermore tends to yield underestimates that encourage active exploration. 

Another source of improvement of CTRACE relative to CE is that updates are made based on a certain measure of confidence for estimates whereas CE updates at every period regardless of confidence. To be more precise on this confidence measure, we first present a high-probability confidence region for least-squares estimates from \cite{Abbasi-YadkoriPalSzepesvari11}. 

\begin{proposition}[Corollary 10 of \cite{Abbasi-YadkoriPalSzepesvari11}] \label{prop:confidenceSet}
\[
\text{Pr} \left( \theta^{*}  \in \mathcal{S}_t (\delta), \,\, \forall t \geq 1 \right) \geq 1 - \delta \quad \text{where} \quad V_t = \kappa I + \sum_{i=1}^{t} \psi_i \psi_i^\top, \quad \hat{\theta}_t = V_t^{-1} \left( \sum_{i=1}^{t} \psi_i \psi_i^\top \theta^{*} +  \sum_{i=1}^{t} \psi_i \epsilon_i \right),
\]
\[
\mathcal{S}_t (\delta) \defeq \left\{ \theta \in \R^{M+1} : (\theta - \hat{\theta}_t )^\top V_t (\theta - \hat{\theta}_t)  
\leq \left( C_\epsilon \sqrt{2\, \textrm{log} \left( \frac{\textrm{det}(V_t)^{1/2} \textrm{det}(\kappa I)^{-1/2} }{\delta} \right)} + \kappa^{1/2} \| \theta_{\text{max}} \| \right)^2  \right\}.
\]
\end{proposition}

This implies that for any $\theta \in \mathcal{S}_t(\delta)$
\[
\| \theta  - \hat{\theta}_t \|^{2} \leq \frac{1}{\lambda_{\textrm{min}}(V_t)} \left( C_\epsilon \sqrt{2\, \log \left( \frac{\textrm{det}(V_t)^{1/2} \textrm{det}(\kappa I)^{-1/2} }{\delta} \right)} + \kappa^{1/2} \| \theta_{\text{max}} \| \right)^2.
\]
By definition, CTRACE updates only when $\lambda_{\textrm{min}}(V_t) \geq \kappa + C_v \, t$. $\lambda_{\textrm{min}}(V_t)$ typically dominates $\log \left( \textrm{det}(V_t) \right)$ for large $t$ because it increases linearly in $t$, and is inversely proportional to the squared estimation error $\| \hat{\theta}_t - \theta^* \|^{2}$. That is, CTRACE updates only when confidence represented by $\lambda_{\textrm{min}}(V_t)$ exceeds the specified level $\kappa + C_v \, t$. From now on, we refer to this updating scheme as confidence-triggered update. Confidence-triggered update makes a significant contribution to reducing the chances of obtaining overestimates of price impact by updating ``carefully'' only at the moments when an upper bound on the estimation error is guaranteed to decrease. 

The minimum inter-update time $\tau \in \mathbb{N}$ in Algorithm \ref{alg:CTRACE} can guarantee that the closed-loop system $\{ z_t \}$ from CTRACE is stable as long as $\tau$ is sufficiently large. Meanwhile, there is no such stability guarantee for CE. The following lemma provides with a specific uniform bound on $\| z_t \|$.
\begin{lemma} \label{lem:uniformBoundCTRACE} 
Under CTRACE with $\tau \geq N \log (2 C_g/\xi) / \log (1/\xi)$ 
\[
\| z_t \| \leq \frac{(2 C_g + 1) C_g C_\omega}{\xi (1-\xi^{\frac{1}{N}})} \defeq C_z^* \quad a.s. \quad \text{and} \quad \| \psi_t \| \leq \frac{(C_g +1)(2 C_g + 1) C_g C_\omega}{\xi (1-\xi^{\frac{1}{N}})} \defeq C_{\psi} \quad a.s. \quad \forall t \geq 0.
\] 
\end{lemma}

Confidence-triggered update yields a good property of CTRACE that CE lacks: CTRACE is {\it inter-temporally consistent} in the sense that estimation errors $\| \theta_t - \theta^* \|$ are bounded with high probability by monotonically nonincreasing upper bounds that converge to zero almost surely as time tends to infinity. The following theorem formally states this property.

\begin{theorem}[Inter-temporal Consistency of CTRACE] \label{thm:intertemporalConsistency}
Let $\{ \theta_t \}$ be estimates generated by CTRACE with $M \geq 2$, $\tau \geq N \log (2 C_g/\xi) / \log (1/\xi)$ and $C_v < \underline{\lambda}_{\psi \psi}^{*}$. Then, the $i$th update time $t_i$ in Algorithm \ref{alg:CTRACE} is finite a.s. Moreover, $\| \theta_t - \theta^{*} \| \leq b_t, \,\, \forall t \geq 0$ on the event $\{ \theta^* \in \mathcal{S}_t(\delta), \,\, \forall t \geq 1 \}$ where  
\[
b_t = \begin{cases} \frac{ 2 C_\epsilon \sqrt{(M+1) \log \left( C_\psi^2 \, t / \kappa + M+1 \right) + 2 \log \left( 1/\delta \right) } + 2\kappa^{1/2} \| \theta_{\text{max}} \| }{ \sqrt{C_v t }} & \text{if } t = t_i \text{ for some } i \\ b_{t-1} & \text{otherwise} \end{cases}, \quad b_0 = \|  \theta_0 - \theta^{*} \|,
\]
and $\{ b_t \}$ is monotonically nonincreasing for all $t \geq 1$ with $\lim_{t \rightarrow \infty} b_t = 0$ a.s. 
\end{theorem}

Moreover, we can show that CTRACE is {\it efficient} in the sense that its $(\epsilon,\delta)$-convergence time is bounded above by a polynomial of $1/\epsilon$, $\log(1/\delta)$ and $\log(1/\delta')$ with probability at least $1-\delta'$. We define {\it $(\epsilon,\delta)$-convergence time} to be the first time when an estimate and all the future estimates following it are within an $\epsilon$-neighborhood of $\theta^{*}$ with probability at least $1 - \delta$. If $\epsilon$ is sufficiently small, we can apply Lemma \ref{lem:persistentExcitationFloating} and Lemma \ref{lem:linearGrowthFloating} to guarantee that $\lambda_{\textrm{min}}(V_t)$ increases linearly in $t$ with high probability after $(\epsilon,\delta)$-convergence time and thereby confidence-triggered update occurs at every $\tau$ periods. This is a critical property that will be used for deriving a poly-logarithmic finite-time expected regret bound for CTRACE. By Theorem \ref{thm:intertemporalConsistency}, it is easy to see that the $(\epsilon,\delta)$-convergence time of CTRACE is bounded above by $t_{N(\epsilon,\delta,C_v)}$ where $N(\epsilon,\delta,C_v)$ is defined as
\[
N(\epsilon,\delta,C_v) = \inf \left\{ i \in \N : \frac{ 2 C_\epsilon \sqrt{(M+1) \log \left( C_\psi^2 \, t_i / \kappa + M+1 \right) + 2 \log \left( 1/\delta \right) } + 2 \kappa^{1/2} \| \theta_{\text{max}} \| }{ \sqrt{ C_v t_i }} \leq \epsilon  \right\}.
\] 
The following theorem presents the polynomial bound on the $(\epsilon,\delta)$-convergence time of CTRACE. 

\begin{theorem}[Efficiency of CTRACE] \label{thm:efficiency}
For any $\epsilon > 0$, $0 < \delta, \delta' < 1$, $\tau \geq N \log (2 C_g/\xi) / \log (1/\xi)$ and $C_v < \frac{7}{8} \underline{\lambda}_{\psi \psi}^{*}$ on the event $\mathcal{B}(\delta')$ defined in Lemma \ref{lem:linearGrowthFixed},
\[
t_{N(\epsilon,\delta,C_v)} \leq T_1^{*}(\delta') \vee \tau + T_2(\epsilon,\delta,C_v) \quad \text{where}
\]
{\small
\[
T_1^{*}(\delta') = 4 \left( \frac{32 (C_z^* C_g)^2 (M+K+1)}{\xi^2 (1-\xi^{\frac{2}{N}}) \underline{\lambda}^{*}_{zz} } \right)^2 \log \left( \frac{(M+K+2)^4}{432 \delta^{'2}} \right) \vee 8 \left( \frac{32 (C_z^* C_g)^2 (M+K+1)}{\xi^2 (1-\xi^{\frac{2}{N}}) \underline{\lambda}^{*}_{zz} } \right)^3 \vee 216,
\]
}
{\footnotesize
\[
T_2(\epsilon,\delta,C_v) = \left( \frac{ 8 C_\epsilon^2 C_\psi (M+1) + 4 \sqrt{ 4 C_\epsilon^4 C_\psi^2 (M+1)^2 + \kappa C_\epsilon^2 C_v \epsilon^2 \left( (M+1)^{3/2} + 2 \log (1/\delta) \right)  } }{ \sqrt{\kappa} C_v \epsilon^2  } \right)^2  \vee \frac{( 4 \kappa \| \theta_{\text{max}} \|)^2}{ C_v \epsilon^2}.
\]
}
\end{theorem}

Finally, we derive a finite-time expected regret bound for CTRACE that is quadratic in logarithm of elapsed time using the efficiency of CTRACE and Lemma \ref{lem:linearGrowthFloating}. 

\begin{theorem}[Finite-Time Expected Regret Bound of CTRACE] \label{thm:finiteTimeBoundCTRACE}
If $\pi$ is CTRACE with $M \geq 2$, $\tau \geq N \log (2 C_g/\xi) / \log (1/\xi)$ and $C_v < \frac{7}{8} \underline{\lambda}_{\psi \psi}^{*}$, then for any $\nu \in (\xi, 1)$ and all $T \geq 2$,
\[
\begin{split}
\bar{R}_T^{\pi}(z_0) &\leq 2 \| P^{*} \| C_z^{*2} + (R + B^\top P^* B) C_z^{*2} C_L^2 \Bigg( \left( \tau_1^{*}(T) + \tau_2^{*}(T) + 1 \right) \| \theta_{\text{max}} \|^2 + \tau_3^{*}(T) \epsilon^{2} \\
& \quad + \frac{ \tau \left( 2 C_\epsilon \sqrt{(M+1) \log \left( C_\psi^2 \, T / \kappa + M+1 \right) + 2 \log \left( 2 T \right) } + 2 \kappa^{1/2} \| \theta_{\text{max}} \| \right)^2 }{ \tilde{C} } \nonumber \\
& \quad \times \log \left( \frac{ \kappa + \tilde{C} (T-1) - (\tilde{C} - C_v)_{+} (\tau_1^{*}(T) + \tau_2^{*}(T))}{ \kappa + \tilde{C} (\tau^{*}(T) - 1) - (\tilde{C} - C_v)_{+} (\tau_1^{*}(T) + \tau_2^{*}(T)) } \right) \1 \{ T > \tau^{*}(T) \} \Bigg)
\end{split}
\]
where $\tilde{C} \defeq \frac{3}{8} \lambda_{\text{min}}(U(\theta^*) \Pi_{zz}(\theta^*) U(\theta^*)^\top )$, $\tau^{*}(T) = \tau_1^{*}(T) + \tau_2^{*}(T) + \tau_3^{*}(T)$,
{\small
\[
\tau_1^{*}(T) = 8 \left( \frac{32 (C_z^* C_g)^2 (M+K+1)}{\xi^2 (1-\xi^{\frac{2}{N}}) \underline{\lambda}^{*}_{zz} } \right)^2 \log \left( \frac{(M+K+2)^2 T}{6 \sqrt{3}} \right) \vee 8 \left( \frac{32 (C_z^* C_g)^2 (M+K+1)}{\xi^2 (1-\xi^{\frac{2}{N}}) \underline{\lambda}^{*}_{zz} } \right)^3 \vee 216 \vee \tau,
\]
}
{\small
\[
\begin{split}
\tau_2^{*}(T) &= \left( \frac{ 8 C_\epsilon^2 C_\psi (M+1) + 4 \sqrt{ 4 C_\epsilon^4 C_\psi^2 (M+1)^2 + \kappa C_\epsilon^2 C_v \epsilon^2 \left( (M+1)^{3/2} + 2 \log (2 T) \right)  } }{ \sqrt{\kappa} C_v \epsilon^2  } \right)^2 \vee  \frac{( 4 \kappa \| \theta_{\text{max}} \|)^2}{ C_v \epsilon^2},
\end{split}
\]
}
{\small
\[
\begin{split}
\tau_3^{*}(T) &= 8 \left( \frac{32 (C_z^* C_g)^2 (M+K+1)}{\xi^2 (1-\xi^{\frac{2}{N}}) \lambda_{\text{min}}(\Pi_{zz}(\theta^*)) } \right)^2 \log \left( \frac{(M+K+2)^2 T}{6 \sqrt{3} } \right) \vee 8 \left( \frac{32 (C_z^* C_g)^2 (M+K+1)}{\xi^2 (1-\xi^{\frac{2}{N}}) \lambda_{\text{min}}(\Pi_{zz}(\theta^*)) } \right)^3 \vee 216 \\
&\quad \vee \frac{3 C_z^* (2 C_\omega + C_z^* )}{C_\omega^2},
\end{split}
\]
}
{\small
\[
\epsilon = \frac{1}{\sqrt{M+1} C_L} \left( \frac{\nu^3 (1-\nu^{\frac{1}{N}})^3 \lambda_{\text{min}}(\Pi_{zz}(\theta^*))}{42 N C_g^{N+1} C_\omega^2} \wedge \frac{\nu^3 (1-\nu^{\frac{1}{N}})^3 \lambda_{\text{min}}(U(\theta^*) \Pi_{zz}(\theta^*) U(\theta^*)^\top)}{42 N C_g^{N+1} C_\omega^2 (1 + \| U(\theta^*) \|)^2} \wedge \frac{\nu - \xi}{N C_g^{N-1}} \right).
\]
}
\end{theorem}

Note that $\tau_1^{*}(T)$, $\tau_2^{*}(T)$ and $\tau_3^{*}(T)$ are all $O(\log T)$. Therefore, it is not difficult to see that the expected regret bound for CTRACE is $O(\log^2 T)$.

\section{Computational Analysis} \label{sec:computationalAnalysis}

In this section, we will compare via Monte Carlo simulation the performance of CTRACE to that of two benchmark policies: CE and a reinforcement learning algorithm recently proposed in \cite{Abbasi-YadkoriSzepesvari10}, which is referred to as AS policy from now on. AS policy was designed to explore efficiently in a broader class of linear-quadratic control problems and appears well-suited for our problem. It updates an estimate only when the determinant of $V_t$ is at least twice as large as the determinant evaluated at the last update, and selects an element from a high-probability confidence region that yields maximum average reward. In our problem, AS policy can translate to update an estimate with $\theta_t = \argmin_{\theta \in \mathcal{S}_t(\delta) \cap \Theta} \text{tr}(P(\theta) \tilde{\Omega})$ at each update time $t$. Intuitively, the smaller price impact,  the larger average profit, equivalently, the smaller $\text{tr}(P(\theta) \tilde{\Omega})$ which is the negative of average profit. In light of this, we restrict our attention to solutions to $\min_{\theta \in \mathcal{S}_t(\delta) \cap \Theta} \text{tr}(P(\theta) \tilde{\Omega})$ of the form $\{ \alpha_t \hat{\theta}_{con,t} \in \mathcal{S}_t(\delta) \cap \Theta : 0 \leq \alpha_t \leq 1 \}$ where $\hat{\theta}_{con,t}$ denotes a constrained least-squares estimate to $\Theta$ with $\ell_2$ regularization. The motivation is to reduce the amount of computation needed for AS policy otherwise it would be prohibitive. Indeed, the minimum appears to be attained always with the smallest $\alpha_t$ such that $\alpha_t \hat{\theta}_{con,t} \in \mathcal{S}_t(\delta) \cap \Theta$, which is provable in the special case considered in Subsection \ref{subsec:closedForm}. Note that $\alpha_t$ can be viewed as a measure of aggressiveness of exploration: $\alpha_t = 1$ means no extra exploration and smaller $\alpha_t$ implies more active exploration. 

\begin{table}[t!] 
\caption{Monte Carlo Simulation Setting (1 trading day = 6.5 hours)} \label{tab:simulationSetting}
\centering
\begin{tabular}{|  p{2.7cm}  |   p{5.8cm}  ||  p{2.9cm}  |  p{3.2cm}  |} 
\hline
$M$ & 6 & $K$ & 2 \\
\hline
Trading interval & 5 mins & Initial asset price & \$50\\
\hline
Half-life of $r$ & [5, 7.5, 10, 15, 30, 45] mins & Half life of factor & [10, 40] mins \\
\hline
$r$ & [ 0.50, 0.63, 0.71, 0.79, 0.89, 0.93 ] & $\Phi$ & diag([0.707, 0.917]) \\
\hline
$\gamma$ (\$/share) & [0, 6, 0, 3, 7, 5] $\times \, 10^{-8}$ & $\lambda$ (\$/share) & $2 \times 10^{-8}$ \\
\hline
$\Sigma_\epsilon$ & $0.0013$ (annualized vol. = 10\%) & $\Omega$ & diag([1, 1]) \\
\hline
$\rho$ & $1 \times 10^{-6}$ & $\theta_{\text{max}}$ & $(5 \times 10^{-7}) \1$ \\
\hline
$\beta$ & $5 \times 10^{-9}$ & $g$ & [0.006, 0.002] \\
\hline
$T$ & $3000$ ($\approx$ 38 trading days) & Sample paths & 600 \\
\hline
\end{tabular}
\end{table}

Table \ref{tab:simulationSetting} summarizes numerical values used in our simulation. The signal-to-noise ratio (SNR), which is defined as $\E[(\lambda u_t + \sum_{m=1}^{M} \gamma_m (d_{m,t} - d_{m,t-1}))^2 ]/\E[\epsilon_t^2]$ under $u_t = L(\theta^*) z_{t-1}$, is 0.058 and the optimal average profit is \$765.19 per period. $\epsilon_t$ and $\omega_t$  are sampled independently from Gaussian distribution even though $\omega_t$ is assumed to be bounded almost surely for the theoretical analysis. In fact, it turns out that the use of Gaussian distribution for $\omega_t$ does not make a noticeable difference from a bounded case. The regularization coefficient $\kappa$, the confidence-triggered update threshold $C_v$, the minimum inter-update time $\tau$ and the significance level $\delta$ are chosen via cross-validation with realized profit: For CTRACE, $\kappa = 1 \times 10^{11}$, $C_v = 600$ and $\tau = 1$. For AS policy, $\kappa = 1 \times 10^8$ and $\delta = 0.99$. The reason for smaller $\kappa$ and large $\delta$ for AS policy is to keep the radius of confidence regions small because the exploration done by AS policy tends to be more than necessary and thus costly. 

\begin{figure}[t!]
\hspace{-1.4cm} 
\includegraphics[scale=0.45]{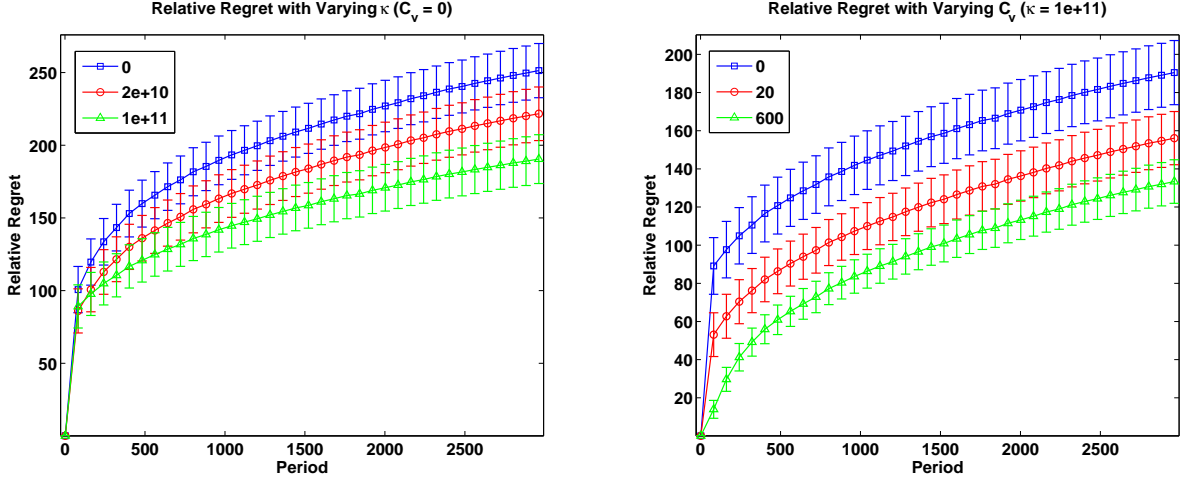}
\caption{Relative regret with varying $\kappa$ and $C_v$: (Left) Varying $\kappa \in \{0, 2 \times 10^{10}, 1 \times 10^{11} \}$ with fixed $C_v = 0$. (Right) Varying $C_v \in \{ 0, 20, 600\}$ with fixed $\kappa = 1 \times 10^{11}$.}
\label{fig:regulConfTrig}
\end{figure}

The left figure in Figure \ref{fig:regulConfTrig} illustrates improvement of relative regret due to regularization. It shows the relative regret of CTRACE with varying $\kappa$ and fixed $C_v = 0$, $\text{i.e.}$ no confidence-triggered update. The vertical bars indicate two standard errors in both directions, that is, approximate 95\% confidence intervals. It is clear that the relative regret is reduced as CTRACE regularizes more, and the improvement from no regularization to $\kappa = 1 \times 10^{11}$ is statistically significant with approximate 95\% confidence level. The right figure in Figure \ref{fig:regulConfTrig} shows improvement achieved by confidence-triggered update with varying $C_v$ but fixed $\kappa = 1 \times 10^{11}$. As you can see, update based on confidence makes a substantial contribution to reducing relative regret further. The improvement from $C_v = 0$ to $C_v = 600$ is statistically significant with approximate 95\% confidence level. 

\begin{figure}[b!]
\hspace{-1.4cm}
\includegraphics[scale=0.45]{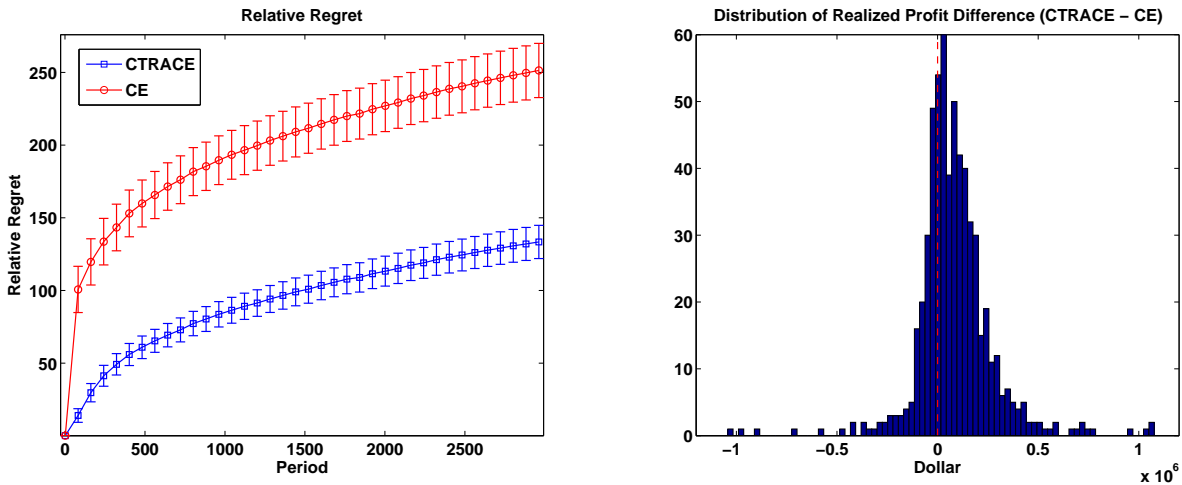}
\caption{(Left) Relative regret of CTRACE and CE. (Right) Distribution of realized profit of CTRACE and CE. The red dotted line represents zero difference. }
\label{fig:CTRACECE}
\end{figure} 

As shown on the left of Figure \ref{fig:CTRACECE}, CTRACE clearly outperforms CE in terms of relative regret and the difference is statistically significant with approximate 95\% confidence level. The dominance stems from both regularization and confidence-triggered update as shown in Figure \ref{fig:regulConfTrig}. The figure on the right shows an empirical distribution of difference between realized profit of CTRACE and that of CE over 600 sample paths. Much more realizations are located to the right with respect to zero profit. It implies that CTRACE tends to make more profit than CE more frequently.

\begin{figure}[t!]
\hspace{-1.4cm}
\includegraphics[scale=0.45]{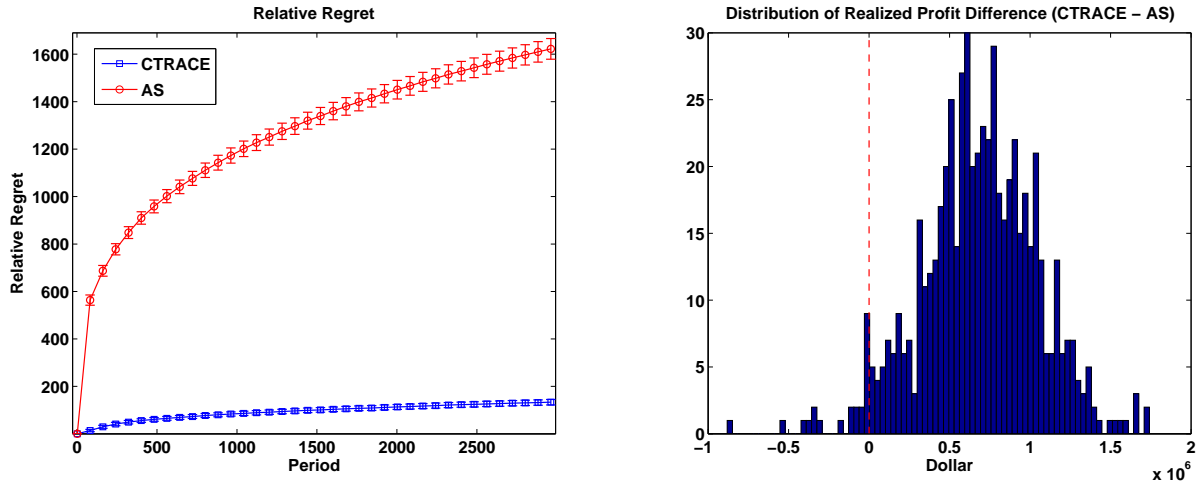}
\caption{(Left) Relative regret of CTRACE and AS policy. (Right) Distribution of realized profit of CTRACE and AS policy. The red dotted line represents zero difference. }
\label{fig:CTRACEAS}
\end{figure}

Finally, we compare performance of CTRACE to that of AS policy in Figure \ref{fig:CTRACEAS}. The left figure shows that CTRACE outperforms AS policy even more drastically than CE in terms of relative regret, and the superiority is statistically significant with approximate 95\% confidence level. On the right, you can see an empirical distribution of difference between realized profit of CTRACE and that of AE over 600 sample paths. It is clear that CTRACE is more profitable than AS policy in most of the sample paths. This illustrates that aggressive exploration performed by AS policy is too costly. The reason is that AS policy is designed to explore actively in situations where pure exploitation done by CE is unable to identify a true model. In our problem, however, a great degree of exploration is naturally induced by observable return-predictive factors and thus aggressiveness of exploration suggested by AS policy turns out to be even more than necessary. Meanwhile, CTRACE strikes a desired balance between exploration and exploitation by taking into account factor-driven natural exploration.

\section{Conclusion} \label{sec:conclusion}

We have considered a dynamic trading problem where a trader maximizes expected average risk-adjusted profit while trading a single security in the presence of unknown price impact. Our problem can be viewed as a special case of reinforcement learning: the trader can improve longer-term performance significantly by making decisions that explore efficiently to learn price impact at the expense of suboptimal short-term behavior such as execution of larger orders than appearing optimal with respect to current information. Like other reinforcement learning problems, it is crucial to strike a balance between exploration and exploitation. To this end, we have proposed the confidence-triggered regularized adaptive certainty equivalent policy (CTRACE) that improves purely exploitative certainty equivalent control (CE) in our problem. The enhancement is attributed to two properties of CTRACE: regularization and confidence-triggered update. Regularization encourages active exploration that accelerates learning as well as reduces the variance of an estimator. It helps keep CTRACE from being a passive learner due to overestimation of price impact that abates trading. Confidence-triggered update allows CTRACE to have monotonically nonincreasing upper bounds on estimation errors so that it reduces the frequency of overestimation. Using these two properties, we derived a finite-time expected regret bound for CTRACE of the form $O(\log^2 T)$. Finally, we have demonstrated through Monte Carlo simulation that CTRACE outperforms CE and a reinforcement learning policy recently proposed in \cite{Abbasi-YadkoriSzepesvari10}. 

As extention to our current model, it would be interesting to develop an efficient reinforcement learning algorithm for a portfolio of securities. Another interesting direction is to incorporate a prior knowledge of particular structures of price impact coefficients, $\text{e.g.}$ sparsity, to an estimation problem. It is worth considering other regularization schemes such as LASSO.

\appendix

\section{Proofs}

\textbf{\textsf{Proof of Theorem \ref{thm:optPolicyExistenceUniqueness} }} Since the evolution of $f_t$ is not affected by $\{ x_t \}$, $\{ d_t \}$ and $\{ u_t \}$, it is not difficult to see that there exists a desired $P$ for our stochastic control problem if there exists $P$ with the same properties for a deterministic control problem having no $f_t$ and $g = 0$. Let $(\tilde{A},\tilde{B},\tilde{Q},\tilde{R},\tilde{S})$ denote reduced coefficient matrices for the deterministic problem of appropriate dimensions. Now, $(\tilde{A},\tilde{B})$ is controllable and this problem is a special case of the problem considered in \cite{Molinari75}. By Theorem 1 in \cite{Molinari75}, there exists a desired $P$ if $\Psi(z) > 0$ for all $z$ on the unit circle where 
\[
\Psi(z) \defeq \left[ \begin{array}{cc} \tilde{B}^\top (I z^{-1} - \tilde{A}^\top)^{-1} & I \end{array} \right]
\left[ \begin{array}{cc} \tilde{Q} & \tilde{S} \\ \tilde{S}^\top & \tilde{R} \end{array} \right] \left[ \begin{array}{c} (I z - \tilde{A})^{-1} \tilde{B} \\ I \end{array} \right].
\]
In our problem, it is not difficult to check that for any $\phi \in (0,2 \pi)$, $\lambda \geq 0$ and $\gamma_i \geq 0$,
\[
\Psi(e^{i \phi}) = \frac{\rho \Sigma_\epsilon}{2(1 - \cos \phi)} + \frac{\lambda}{2} + \sum_{m=1}^{M} \frac{2 \gamma_m (1 - r_m \cos \phi )}{1 + r_m^2 - 2 r_m \cos \phi} > 0
\]
and $\lim_{\phi \rightarrow 0} \Psi(e^{i \phi}) = \infty > 0$. Therefore, the desired result follows. Noting an upper block diagonal structure of the original closed-loop system matrix $A + B L$, we can easily see that the stability for the deterministic problem carries over to our original problem. The uniqueness of a stabilizing solution follows from the stability. For the optimality of $\pi$, we can use the same proof in Chapter 4 of \cite{Bertsekas05}. \qed  

\textbf{\textsf{Proof of Lemma \ref{lem:uniformBound} }} By Theorem \ref{thm:optPolicyExistenceUniqueness}, $\rho_{\text{sr}}(G(\theta)) < 1$ for all $\theta \in \Theta$.
Since $\Theta$ is a compact set and Assumption \ref{assum:optimalPolicy}-(a) implies the continuity of $L(\theta)$ and $G(\theta)$, it follows that $\sup_{\theta \in \Theta} \| G(\theta) \| < \infty$ and $\sup_{\theta \in \Theta} \rho_{\text{sr}}(G(\theta)) < 1$. Therefore, by Theorem in \cite{BuchananParlett66}, $\{ G^n(\theta) \}$ uniformly converges to zero matrix. That is, for any $0 < \xi < 1$, there exists $N \in \mathbb{N}$ being independent of $\theta$ such that $\| G^N(\theta) \| \leq \xi$ for all $\theta \in \Theta$. Also, $\max_{0 \leq i \leq N-1} \sup_{\theta \in \Theta} \| G^i(\theta) \| < \infty$ by continuity of $G(\theta)$ and compactness of $\Theta$. For any $t \geq 0$, it is easy to see that $\| G^t(\theta) \| \leq C_g \xi^{\lfloor t/N \rfloor}$ by definition of $C_g$ and $N$. Since $z_t = G^t(\theta) z_0 + \sum_{i=1}^{t} G^{t-i}(\theta) W_t$,
\[
\begin{split}
\| z_t \| &\leq \| G^t(\theta) \| \| z_0 \| + \sum_{i=1}^{t} \| G^{t-i}(\theta)) \| \| W_t \| \leq C_g \xi^{\lfloor t/N \rfloor} \| z_0 \| + \sum_{i=1}^{t} C_g \xi^{\lfloor (t-i)/N \rfloor} C_\omega \\
&\leq C_g \| z_0 \| + C_g C_\omega \sum_{i=1}^{t} \xi^{(t-i)/N-1} \leq C_g \| z_0 \| + C_g C_\omega / (\xi (1 - \xi^{1/N})) \,\, a.s.
\end{split}
\]
Since $U(\theta) = ( G(\theta) )_{1:M+1,*} - [I \,\,\, 0]$, it follows that $\| U(\theta) \| \leq \| (G(\theta) )_{1:M+1,*} \| + \| [ I \,\,\, 0] \| \leq C_g + 1$. \qed

\textbf{\textsf{Proof of Lemma \ref{lem:persistentExcitationFixed} }} For notational simplicity, let $G = G(\theta)$, $L = L(\theta)$ and $\Pi_{zz} = \Pi_{zz}(\theta)$. The almost-sure convergence in (\ref{eqn:ergodic}) follows from Lemma 2 in \cite{AndersonTaylor79}. It is easy to see that $U(\theta)$ is full-rank since $(L)_1 \neq 0$. Therefore, it is sufficient to show that $\Pi_{zz}$ is positive definite. Since $G$ is a stable matrix and $\tilde{\Omega} \succeq 0$, $\Pi_{zz} = \sum_{i=0}^{\infty} G^i \tilde{\Omega} (G^\top)^i \succeq \sum_{i=0}^{M+K} G^i \tilde{\Omega} (G^\top)^i = H H^\top$ where $H = \left[  \tilde{\Omega}^{1/2} \,\, G \tilde{\Omega}^{1/2} \,\, \ldots \,\, G^{M+K} \tilde{\Omega}^{1/2} \right]$.
Thus, it is sufficient to show that $H$ is full-rank. First, we will show that $\{ (G)_{1:M+1,M+2}, \ldots, (G^{M+1})_{1:M+1,M+2} \}$ is linearly independent. We can show by induction that
$(G^i)_{*,M+2} = \left[ g_i(1) \,\,\, g_i(r_1) \,\,\, \cdots \,\,\, g_i(r_M) \,\,\, h_i  \right]^\top$ where $g_i (r) = (L)_{M+2} \sum_{m=0}^{i-1} (\Phi^{m})_{1,1} r^{i-1-m}$ and $h_i = (\Phi^i)_{*,1}$.
Since each $g_i(r)$ is a polynomial of degree $i-1$ and its leading coefficient is all $(L)_{M+2} \neq 0$, we can transform $[ (G)_{1:M+1,M+2}, \,\, \ldots \,\, (G^{M+1})_{1:M+1,M+2} ]$ into Vandermonde matrix through elementary row operations. Thus, $[ (G)_{1:M+1,M+2} \,\, \ldots \,\, (G^{M+1})_{1:M+1,M+2} ]$ is nonsingluar. 
Now, suppose $\alpha^\top H = 0$ for some $\alpha \in \R^{M+K+1}$. By definition of $H$ and $\tilde{\Omega}$, it implies $(\alpha)_{M+2:M+K+1} = 0$. Then, by nonsingularity of $[ (G)_{1:M+1,M+2} \,\, \ldots \,\, (G^{M+1})_{1:M+1,M+2} ]$, we may conclude $\alpha_{1:M+1}^\top = 0$. Therefore, $\alpha = 0$ and we may conclude that $H$ is full-rank. \qed 

\textbf{\textsf{Proof of Corollary \ref{cor:extremums} }} By Assumption \ref{assum:optimalPolicy}-(a), $L(\theta)$ is continuous on $\Theta$ and so are $G(\theta)$ and $U(\theta)$. Uniform convergence of $\E \left[ \frac{1}{T} \sum_{t=1}^{T} z_{t-1} z_{t-1}^\top \right]$ to $\Pi_{zz}(\theta)$ on $\Theta$ follows from the fact that for any $\epsilon > 0$
\begin{align*}
&\left\| \E\left[ \frac{1}{T} \sum_{t=1}^{T} z_{t-1} z_{t-1}^\top \right] - \Pi_{zz}(\theta) \right\| = \left\| \frac{1}{T} z_{0} z_{0}^\top + \sum_{t=1}^{T-1} \frac{t-1}{T} G^{t-1} \tilde{\Omega} (G^\top)^{t-1} + \sum_{t=T}^{\infty} G^{t-1} \tilde{\Omega} (G^\top)^{t-1} \right\| \\
& \leq \frac{\| z_0 \|^2}{T} + \left\| \tilde{\Omega} \right\| \frac{C_g^2}{\xi^2} \left( \frac{1}{T} \frac{\xi^\frac{2}{N}}{(1 - \xi^\frac{2}{N})^2} + \frac{\xi^\frac{2(T-1)}{N}}{1-\xi^\frac{2}{N}} \right) \leq \epsilon \quad \text{for sufficiently large } T \text{ independent of } \theta.
\end{align*}
Since $\E\left[ \frac{1}{T} \sum_{t=1}^{T} z_{t-1} z_{t-1}^\top \right] = \frac{1}{T} z_0 z_0^\top + \frac{1}{T} \sum_{t=1}^{T-1} \sum_{i=0}^{t-1} G^i \tilde{\Omega} (G^\top)^i$ is continuous in $\theta \in \Theta$ for all $T \geq 1$, the limiting matrix $\Pi_{zz}(\theta)$ is continuous in $\theta \in \Theta$ component-wise. Thus, so is $U(\theta) \Pi_{zz}(\theta) U(\theta)^\top$. Finally, $\lambda_{\text{min}} \left( U(\theta) \Pi_{zz}(\theta) U(\theta)^\top\right)$ is continuous on $\Theta$. Since $\lambda_{\text{min}} \left( U(\theta) \Pi_{zz}(\theta) U(\theta)^\top\right) > 0, \,\, \forall \theta \in \Theta$ and $\Theta$ is a compact set, it follows from its continuity that $\inf_{\theta \in \Theta} \lambda_{\text{min}} \left( U(\theta) \Pi_{zz}(\theta) U(\theta)^\top\right) > 0$. \qed 

\textbf{\textsf{Proof of Lemma \ref{lem:linearGrowthFixed} }} Let $e_i \in \R^{M+K+1}$ denote an elementary vector whose entries are all zero except for $i$th entry being one and $\eta_{ij,k} \defeq e_i^\top z_k z_k^\top e_j - e_i^\top \E[ z_k z_k^\top | \mathcal{F}_{k-1} ] e_j, \,\, 1 \leq i, j, \leq M+K+1$. Since $| \eta_{ij,k} | \leq 2 C_z^2 \,\, a.s.$, $\{ \eta_{ij,k} \}$ is an almost-surely bounded martingale difference process adapted to $\{ \mathcal{F}_k \}$ and thus it is conditionally sub-Gaussian with $\E[ \exp(\gamma \eta_{ij,k}) | \mathcal{F}_{k-1} ] \leq \exp \left( \gamma^2 (2 C_z^2)^{2} / 2 \right) \,\, a.s.$ Hence, if we use a special case of Corollary 1 in \cite{Abbasi-YadkoriPalSzepesvari11} with $m_k = 1$ for all $k$, then for all $1 \leq i, j \leq M+K+1$ and any $a > 0$
\[
\text{Pr} \left( \left| \sum_{k=1}^{t} \eta_{ij,k} \right| \leq 2 C_z^2 \sqrt{ (a+t) \log \left( \frac{(M+K+2)^4 (a+t)}{4 a \delta^2} \right)} \quad \forall t \geq 1 \right) \geq 1 - \frac{2\delta}{(M+K+2)^2}.
\]
Using $\eta_{ij,k} = \eta_{ji,k}$ and $\E[ z_k z_k^\top | \mathcal{F}_{k-1} ]  = G(\theta) z_{k-1} z_{k-1}^\top G(\theta)^\top + \tilde{\Omega}$, it follows from the union bound that $\text{Pr} \Bigg( | \left( Y_t \right)_{ij} | \leq \epsilon, \,\, 1 \leq i, j \leq M+K+1, \,\, \forall t \geq t^*(\delta,\epsilon,a) \Bigg) \geq 1 - \delta$
where $Y_t \defeq \frac{1}{t} \sum_{k=1}^{t} z_{k} z_{k}^\top - G(\theta) \left( \frac{1}{t} \sum_{k=1}^{t} z_{k-1} z_{k-1}^\top \right) G(\theta)^\top  - \tilde{\Omega}$ and
\[
t^*(\delta,\epsilon,a) \defeq 4 \left( \frac{2 C_z^2}{\epsilon} \right)^2 \log \left( \frac{(M+K+2)^4 }{2 a \delta^2} \right) \linebreak \vee 8 \left( \frac{2 C_z^2}{\epsilon} \right)^3 \vee a \vee 216.
\]
On the above event, $\| Y_t \| \leq \| Y_t \|_F \leq (M+K+1) \epsilon$ and $-(M+K+1) \epsilon I \preceq Y_t \preceq (M+K+1) \epsilon I, \,\, \forall t \geq t^*(\delta,\epsilon,a)$. We can rewrite $-(M+K+1) \epsilon I \preceq Y_t$ as
\[
\frac{1}{t} \sum_{k=1}^{t} z_{k-1} z_{k-1}^\top \succeq G(\theta) \left( \frac{1}{t} \sum_{k=1}^{t} z_{k-1} z_{k-1}^\top \right) G(\theta)^\top + \tilde{\Omega} - (M+K+1) \epsilon I + \left( \frac{1}{t} z_{0} z_{0}^\top -  \frac{1}{t} z_{t} z_{t}^\top \right).
\]
Repeating $n$ times a process of left-multiplying both sides with $G(\theta)$, right-multiplying with $G(\theta)^\top$ and adding the resulting inequality into the original one side-by-side, we obtain   
\[
\begin{split}
\frac{1}{t} \sum_{k=1}^{t} z_{k-1} z_{k-1}^\top &\succeq G^{n+1}(\theta) \left( \frac{1}{t} \sum_{k=1}^{t} z_{k-1} z_{k-1}^\top \right) G^{n+1}(\theta)^\top + \sum_{i=0}^{n} G^{i}(\theta) \tilde{\Omega} G^{i}(\theta)^\top \\
& \quad - (M+K+1) \epsilon \sum_{i=0}^{n} G^{i}(\theta) G^{i}(\theta)^\top + \sum_{i=0}^{n} G^{i}(\theta) \left( \frac{1}{t} z_{0} z_{0}^\top -  \frac{1}{t} z_{t} z_{t}^\top \right) G^{i}(\theta)^\top.
\end{split}
\]
Note that $\left\| \sum_{i=0}^{n} G^{i}(\theta) \left( \frac{1}{t} z_{0} z_{0}^\top -  \frac{1}{t} z_{t} z_{t}^\top \right) G^{i}(\theta)^\top \right\| \leq \sum_{i=0}^{n}  \frac{2}{t} C_z^2 \| G^{i}(\theta) \|^{2} \leq 2 C_z^2 C_g^2 / (t \xi^2 (1 - \xi^{2/N}))$ and $\left\| \sum_{i=0}^{n} G^{i}(\theta) G^{i}(\theta)^\top \right\| \leq \sum_{i=0}^{n}  \| G^{i}(\theta) \|^{2} \leq C_g^2 / (\xi^2 (1 - \xi^{2/N}))$.
Taking limit over $n$ and using these two inequalities, we have with probability at least $1-\delta$
\[
\frac{1}{t} \sum_{k=1}^{t} z_{k-1} z_{k-1}^\top \succeq \Pi_{zz}(\theta) - \left( \frac{C_g^2 (M+K+1)}{\xi^2 (1 - \xi^{\frac{2}{N}})}\epsilon + \frac{1}{t} \frac{2 C_z^2 C_g^2}{\xi^2 (1 - \xi^{\frac{2}{N}})}  \right) I, \quad \forall t \geq t^*(\delta,\epsilon,a).
\]
Setting $\epsilon = \xi^2 (1 - \xi^{2/N}) \lambda_{\text{min}}(\Pi_{zz}(\theta))/(16 C_g^2 (M+K+1))$ and $a = 216$, we have $\frac{1}{t} \sum_{k=1}^{t} z_{k-1} z_{k-1}^\top \succeq \Pi_{zz}(\theta) - \frac{\lambda_{\text{min}}(\Pi_{zz}(\theta))}{8} I$ for all $t \geq t^*(\delta,\epsilon,a) \vee 32 C_z^2 C_g^2 / (\xi^2 (1-\xi^{2/N}) \lambda_{\text{min}}(\Pi_{zz}(\theta)))$.
It is easy to show that
$
t^*(\delta,\epsilon,a) \geq 32 C_z^2 C_g^2 / (\xi^2 (1-\xi^{2/N}) \lambda_{\text{min}}(\Pi_{zz}(\theta))).
$
Similarly, from $Y_t \preceq (M+K+1) \epsilon I$, we can obtain for all $t \geq t^*(\delta,\epsilon,a)$
\[
\frac{1}{t} \sum_{k=1}^{t} z_{k-1} z_{k-1}^\top \preceq \Pi_{zz}(\theta) + \left( \frac{(M+K+1) C_g^2}{\xi^2 (1 - \xi^{\frac{2}{N}})}\epsilon - \frac{1}{t} \frac{2 C_z^2 C_g^2}{\xi^2 (1 - \xi^{\frac{2}{N}})}  \right) I \preceq \Pi_{zz}(\theta) + \frac{ \lambda_{\text{min}}(\Pi_{zz}(\theta))}{16} I.  
\]
Since $\lambda_{\text{min}}(\Pi_{zz}(\theta)) I \preceq \Pi_{zz}(\theta)$, it follows that $\frac{7}{8} \Pi_{zz}(\theta) \preceq \frac{1}{t} \sum_{k=1}^{t} z_{k-1} z_{k-1}^\top \preceq \frac{17}{16} \Pi_{zz}(\theta)$ and thus $\frac{7}{8} U(\theta) \Pi_{zz}(\theta) U(\theta)^\top \preceq U(\theta) \frac{1}{t} \sum_{k=1}^{t} z_{k-1} z_{k-1}^\top U(\theta)^\top \preceq \frac{17}{16} U(\theta) \Pi_{zz}(\theta) U(\theta)^\top$. \qed

\textbf{\textsf{Proof of Lemma \ref{lem:persistentExcitationFloating} }} For notational convenience, let $G = G(\theta)$, $G_t = G(\theta_t)$, $U = U(\theta)$, $U_t = U(\theta_t)$, $\Pi_{zz} = \Pi_{zz}(\theta)$ and $\Pi(i,j) = G_i \cdots G_j$. By definition of $G$ and $\eta$, $\| G_t - G \| \leq \| B \| \| L(\theta_t) - L(\theta) \| \leq \sqrt{M+1} \, C_L \| \theta_t - \theta \| \leq \eta, \,\, \forall t \geq 0.$ Since $z_t$ can be expressed as $z_t = \Pi(0,t-1) z_0 + \sum_{i=1}^{t} \Pi(i,t-1) W_i$, we have
\[
\begin{split}
\frac{1}{T} \sum_{t=1}^{T} z_{t-1} z_{t-1}^\top &= \frac{1}{T} \sum_{t=1}^{T} \left( G^{t-1} z_0 + \sum_{i=1}^{t-1} G^{t-i-1} W_i \right) \left(G^{t-1} z_0 + \sum_{j=1}^{t-1} G^{t-j-1} W_j \right)^\top \\
& \quad + \frac{1}{T} \sum_{t=1}^{T} \left( \Pi(0,t-2) z_0 z_0^\top \Pi(0,t-2)^\top - G^{t-1} z_0 z_0^\top G^{t-1 \top} \right) \quad \cdots \,\, (a) \nonumber  \\
& \quad + \frac{1}{T} \sum_{t=2}^{T} \sum_{j=1}^{t-1} \left( \Pi(0,t-2) z_0 W_j^\top \Pi(j,t-2)^\top - G^{t-1} z_0 W_j^\top G^{t-j-1 \top} \right) \quad \cdots \,\, (b) \nonumber  \\
& \quad + \frac{1}{T} \sum_{t=2}^{T} \sum_{j=1}^{t-1} \left( \Pi(j,t-2) W_j z_0^\top \Pi(0,t-2)^\top - G^{t-j-1} W_j z_0^\top G^{t-1 \top} \right) \quad \cdots \,\, (c) \nonumber  \\
& \quad + \frac{1}{T} \sum_{t=2}^{T} \sum_{i=1}^{t-1} \sum_{j=1}^{t-1} \left( \Pi(i,t-2) W_i W_j^\top \Pi(j,t-2)^\top - G^{t-i-1} W_i W_j^\top G^{t-j-1 \top} \right)  \quad \cdots \,\, (d) \nonumber 
\end{split}
\]
Then, we can show that
\[
\| (a) \| \leq \frac{9 \eta N C_g^{N+1} \| z_0 \|^2 }{ T \nu^3 (1 - \nu^{2/N})^2}, \quad \| (b) \|, \| (c) \| \leq \frac{9 \eta N C_g^{N+1} C_\omega \| z_0 \|}{T \nu^3 (1 - \nu^{1/N})^3} \quad \| (d) \| \leq \frac{18 \eta N C_g^{N+1} C_\omega^2}{ \nu^3 (1 - \nu^{1/N})^3} \quad \text{and}
\]
\[
\| (a) + (b) + (c) + (d) \| \leq \frac{21 \eta N C_g^{N+1} C_\omega^2 }{ \nu^3 (1 - \nu^{1/N})^3} \leq \frac{\lambda_{\text{min}}(\Pi_{zz})}{2}, \quad \forall T \geq \frac{3 \| z_0 \| (2 C_\omega + \| z_0 \|)}{C_\omega^2}.
\]
It follows that
\[
\frac{1}{T} \sum_{t=1}^{T} z_{t-1} z_{t-1}^\top  \succeq 
\frac{1}{T} \sum_{t=1}^{T} \left( G^{t-1} z_0 + \sum_{i=1}^{t-1} G^{t-i-1} W_i \right) \left(G^{t-1} z_0 + \sum_{j=1}^{t-1} G^{t-j-1} W_j \right)^\top
- \frac{\lambda_{\text{min}}(\Pi_{zz})}{2} I.
\]
Taking $\liminf$ on both sides, $\liminf_{T \rightarrow \infty} \frac{1}{T} \sum_{t=1}^{T} z_{t-1} z_{t-1}^\top \succeq \Pi_{zz} - \frac{\lambda_{\text{min}}(\Pi_{zz})}{2} I \succeq \frac{\lambda_{\text{min}}(\Pi_{zz})}{2} I \,\, a.s.$ Likewise, we can show that $\liminf_{T \rightarrow \infty} \frac{1}{T} \sum_{t=1}^{T} \psi_t \psi_t^\top \succeq U \Pi_{zz} U^\top - \frac{\lambda_{\text{min}}(U \Pi_{zz} U^\top)}{2} I \succeq \frac{\lambda_{\text{min}}(U \Pi_{zz} U^\top)}{2} I \,\, a.s.$ \qed

\textbf{\textsf{Proof of Lemma \ref{lem:linearGrowthFloating} }} Using the same techniques in the proof of Lemma \ref{lem:linearGrowthFixed} and Lemma \ref{lem:persistentExcitationFloating}, we can obtain that on the event $\mathcal{B}(\delta)$ with $\text{Pr}(\mathcal{B}(\delta)) \geq 1 - \delta$, 
\begin{align*}
\frac{1}{T} \sum_{t=1}^{T} \psi_t \psi_t^\top & \succeq U(\theta) \frac{1}{T} \sum_{t=1}^{T} \left( G(\theta)^{t-1} z_0 + \sum_{i=1}^{t-1} G(\theta)^{t-i-1} W_i \right) \left(G(\theta)^{t-1} z_0 + \sum_{j=1}^{t-1} G(\theta)^{t-j-1} W_j \right)^\top U(\theta)^\top \\
& \quad - \frac{\lambda_{\text{min}}(U(\theta) \Pi_{zz}(\theta) U(\theta)^\top)}{2} I  \\
& \succeq U(\theta) \left( \Pi_{zz}(\theta) - \frac{\lambda_{\text{min}}(\Pi_{zz}(\theta))}{8} I \right) U(\theta)^\top - \frac{\lambda_{\text{min}}(U(\theta) \Pi_{zz}(\theta) U(\theta)^\top)}{2} I   \\
& \succeq \frac{3}{8} \lambda_{\text{min}}(U(\theta) \Pi_{zz}(\theta) U(\theta)^\top ), \quad \forall T \geq T_1(\| z_0 \|, \theta, \delta) \vee \frac{3 \| z_0 \| (2 C_\omega + \| z_0 \|)}{C_\omega^2}. \qed 
\end{align*}

\textbf{\textsf{Proof of Lemma \ref{lem:uniformBoundCTRACE} }} 
Using $\| z_{t_i + j} \| \leq C_g \xi^{j/N-1} \| z_{t_i} \| + C_g C_\omega / (\xi (1 - \xi^{1/N} ) ) \,\, a.s.$ for $j \leq t_{i+1} - t_i$ and $C_g \xi^{\tau/N - 1} \leq \frac{1}{2}$, we can show by induction that $\| z_{t_i} \| \leq 2 C_g C_\omega / (\xi (1-\xi^{1/N})) \,\, a.s.$ for all $i \geq 1$. For any $t_{i} < t < t_{i+1}$, $\| z_t \| \leq C_g \xi^{(t-t_i)/N-1} \| z_{t_i} \| + C_g C_\omega / (\xi (1-\xi^{1/N})) \leq C_z^* \,\, a.s.$ Finally, $\| \psi_t \| \leq \| U(\theta_{t-1}) \| \| z_{t-1} \| \leq (C_g + 1) C_z^* = C_\psi$. \qed 

\textbf{\textsf{Proof of Theorem \ref{thm:intertemporalConsistency} }} Given
$C_v < \underline{\lambda}_{\psi \psi}^{*} \leq \lambda_{\text{min}}\left( U(\theta) \Pi_{zz}(\theta) U(\theta)^\top \right)$, it is easy to show that  $\text{Pr}(t_i < \infty, \,\, \forall i \geq 1) = 1$. Using 
$\theta_{t_i} = \argmin_{\theta \in \Theta} \,\, \sum_{j=1}^{t_i} \left( (\Delta p_j - g^\top f_{j-1}) - \psi_j^\top \theta  \right)^2 + \kappa \| \theta \|^2 = \argmin_{\theta \in \Theta} \,\, (\theta - \hat{\theta}_{t_i})^\top V_{t_i} (\theta - \hat{\theta}_{t_i})$ and Proposition \ref{prop:confidenceSet}, we can show that on the event $\{ \theta^* \in \mathcal{S}_t(\delta) \,\, \forall t \geq 1\}$ for any $i \geq 1$
{\small
\begin{align*}
\| \theta_{t_i} - \theta^{*} \| \leq \| \theta_{t_i} - \hat{\theta}_{t_i} \| + \| \hat{\theta}_{t_i} - \theta^{*} \| \leq \frac{2 C_\epsilon \sqrt{(M+1) \log \left( C_\psi^2 \, t_i / \kappa + M+1 \right) + 2 \log \left( 1/\delta \right) } + 2 \kappa^{1/2} \| \theta_{\text{max}} \| }{ \sqrt{ C_v t_i}} = b_{t_i}.
\end{align*}
}
For any $t_i < t < t_{i+1}$, $\| \theta_{t} - \theta^{*} \| = \| \theta_{t_i} - \theta^{*} \| \leq b_{t_i} = b_t$. It is easy to show through elementary calculus that $b_{t_i}$ is strictly decreasing in $t_i \geq 1$ if $M \geq 2$. \qed

\textbf{\textsf{Proof of Theorem \ref{thm:efficiency} }}
Using $\log(t+M+1) \leq \sqrt{t} + \sqrt{M+1}$ for all $t \geq 0$, we can show that 
\[
\frac{ 2 C_\epsilon \sqrt{(M+1) \log \left( C_\psi^2 \, t / \kappa + M+1 \right) + 2 \log \left( 1/\delta \right) } + 2 \kappa^{1/2} \| \theta_{\text{max}} \| }{ \sqrt{ C_v t }} \leq \epsilon, \quad \forall t \geq T_2(\epsilon,\delta,C_v).
\]
Suppose for contradiction that $t_{N(\epsilon,\delta,C_v)} > T_1^{*}(\delta') \vee \tau + T_2(\epsilon,\delta,C_v) \defeq \tilde{T}^{*}$. Let $t_i$ be the last update time less than $T_2(\epsilon,\delta,C_v)$. Then, there is no update time in the interval $[t_i+1, \tilde{T}^{*}]$ by definition of $t_{N(\epsilon,\delta,C_v)}$ and $T_2(\epsilon,\delta,C_v)$. By definition of $t_i$ and Lemma \ref{lem:linearGrowthFixed}, 
\[
\lambda_{\text{min}} \left( V_{\tilde{T}^{*}} \right) \geq \lambda_{\text{min}}( V_{t_i} ) + \lambda_{\text{min}} \left( \sum_{t = t_i + 1}^{\tilde{T}^{*}} \psi_t \psi_t^\top \right) \geq \kappa + C_v t_i +  \frac{7}{8} \underline{\lambda}_{\psi \psi}^{*} ( \tilde{T}^{*} - t_i) \geq \kappa + C_v \tilde{T}^{*}. 
\]
It is clear that $\tilde{T}^{*} - t_i \geq \tau$. Consequently, $\tilde{T}^{*}$ is eligible for a next update time after $t_i$. It implies that $t_{N(\epsilon,\delta,C_v)} = \tilde{T}^{*}$ but this is a contradiction. \qed 

\textbf{\textsf{Proof of Theorem \ref{thm:finiteTimeBoundCTRACE} }} Note that
\[
(R + B^\top P^* B) \sum_{t=1}^{T} ( (L(\theta_{t-1}) - L(\theta^{*})) z_{t-1} )^2 \leq (R + B^\top P^* B) C_z^{*2}  C_L^2 \sum_{t=1}^{T}  \| \theta_{t-1} - \theta^{*} \|^2.
\]
Set $\delta = 1/T$. Then, on the event $\mathcal{A}(T) \defeq \{ \theta^{*} \in \mathcal{S}_t(1/(2 T)) \,\,\, \forall t \geq 1\} \cap \mathcal{B}(1/(2 T))$ with $\text{Pr} \left( \mathcal{A}(T) \right) \geq 1 - 1/T$, we have
\[
\sum_{t=1}^{\tau_1^{*}(T) + \tau_2^{*}(T)} \| \theta_{t-1} - \theta^* \|^2 \leq (\tau_1^{*}(T) + \tau_2^{*}(T)) \| \theta_{\text{max}} \|^2, \quad \sum_{t=\tau_1^{*}(T) + \tau_2^{*}(T)+1}^{\tau^{*}(T)} \| \theta_{t-1} - \theta^* \|^2 \leq \tau_3^{*}(T) \epsilon^{2}.
\]
By Lemma \ref{lem:linearGrowthFloating},  
\[
\begin{split}
\lambda_{\text{min}} \left( V_{t-1} \right) &\geq \lambda_{\text{min}} \left( V_{t_{N(\epsilon,1/(2 T),C_v)}} \right) + \lambda_{\text{min}} \left( \sum_{i=t_{N(\epsilon,1/(2 T),C_v)}+1}^{t-1} \psi_i \psi_i^\top \right) \\
& \geq \kappa + \tilde{C} (t-1) - (\tilde{C} - C_v)_{+} (\tau_1^{*}(T) + \tau_2^{*}(T)), \quad \forall t \geq \tau^{*}(T).
\end{split}
\]
Therefore,
{\small 
\begin{align*}
&\sum_{t=\tau^{*}(T)+1}^{T} \| \theta_{t-1} - \theta^* \|^2 \leq \sum_{t=\tau^{*}(T)+1}^{T} \frac{ \tau}{\lambda_{\textrm{min}}(V_{t-1})} \left( 2 C_\epsilon \sqrt{2\, \log \left( \frac{\textrm{det}(V_{t-1})^{1/2} \textrm{det}(\kappa I)^{-1/2} }{\delta/2} \right)} + 2 \kappa^{1/2} \| \theta_{\text{max}} \| \right)^2 \\
& \leq \frac{ \tau \left( 2 C_\epsilon \sqrt{(M+1) \log \left( C_\psi^2 \, T / \kappa + M+1 \right) + 2 \log \left( 2 T \right) } + 2 \kappa^{1/2} \| \theta_{\text{max}} \| \right)^2 }{ \tilde{C} } \\
& \quad \times \log \left( \frac{ \kappa + \tilde{C} (T-1) - (\tilde{C} - C_v)_{+} (\tau_1^{*}(T) + \tau_2^{*}(T)) }{ \kappa + \tilde{C} (\tau^{*}(T) - 1) - (\tilde{C} - C_v)_{+} (\tau_1^{*}(T) + \tau_2^{*}(T)) } \right).
\end{align*}
}
Let $q = \text{Pr} \left( \mathcal{A}(T) \right)$, $L_{t} = L(\theta_{t})$ and $L^* = L(\theta^{*})$. Then,
\begin{align*}
\bar{R}_T^{\pi} (z_0) &= \E[ z_T^{* \top} P^* z_T^{*}  - z_T^\top P^* z_T ] + \E \left[ \sum_{t=1}^{T} (L_{t-1} z_{t-1} - L^{*} z_{t-1})^\top (R + B^\top P^* B) (L_{t-1} z_{t-1} - L^{*} z_{t-1}) \right] \\
& \leq 2 \| P^{*} \| C_z^2 + q \E \left[ \left. \sum_{t=1}^{T} (L_{t-1} z_{t-1} - L^{*} z_{t-1})^\top (R + B^\top P^* B) (L_{t-1} z_{t-1} - L^{*} z_{t-1}) \right| \mathcal{A}(T) \right] \\
& \quad + (1-q) \E \left[ \left. \sum_{t=1}^{T} (L_{t-1} z_{t-1} - L^{*} z_{t-1})^\top (R + B^\top P^* B) (L_{t-1} z_{t-1} - L^{*} z_{t-1}) \right| \mathcal{A}(T)^c \right] \\
& \leq 2 \| P^{*} \| C_z^2 + \E \left[ \left. \sum_{t=1}^{T} (L_{t-1} z_{t-1} - L^{*} z_{t-1})^\top (R + B^\top P^* B) (L_{t-1} z_{t-1} - L^{*} z_{t-1}) \right| \mathcal{A}(T) \right] \\
& \quad + (R + B^\top P^* B) C_z^2 C_L^2 \| \theta_{\text{max}} \|^2 \quad \left( \because \,\, 1 - \frac{1}{T} \leq q \leq 1 \right) \\
& \leq  2 \| P^{*} \| C_z^2 + (R + B^\top P^* B) C_z^2 C_L^2 \Bigg( \left( \tau_1^{*}(T) + \tau_2^{*}(T) + 1 \right) \| \theta_{\text{max}} \|^2 + \tau_3^{*}(T) \epsilon^{2} \\
& \quad + \frac{ \tau \left( 2 C_\epsilon \sqrt{(M+1) \log \left( C_\psi^2 \, T / \kappa + M+1 \right) + 2 \log \left( 2 T \right) } + 2 \kappa^{1/2} \| \theta_{\text{max}} \| \right)^2 }{ \tilde{C} } \\
& \quad \times \log \left( \frac{ \kappa + \tilde{C} (T-1) - (\tilde{C} - C_v)_{+} (\tau_1^{*}(T) + \tau_2^{*}(T)) }{ \kappa + \tilde{C} ( \tau^{*}(T) - 1) - (\tilde{C} - C_v)_{+} (\tau_1^{*}(T) + \tau_2^{*}(T)) } \right) \1 \left\{ T >  \tau^{*}(T) \right\} \Bigg). \qed
\end{align*}

\bibliography{adaptive_execution}

\end{document}